\documentclass[aps,prl,twocolumn,superscriptaddress,10pt]{revtex4-2}
\usepackage{multirow,amsthm,amssymb,amsbsy,amsmath,epsfig,epstopdf,bm,float}
\DeclareMathOperator{\Tr}{Tr}
\usepackage{enumitem}
\usepackage{graphicx}
\usepackage{booktabs}
\usepackage{verbatim}
\usepackage{dcolumn}
\usepackage{color}
\usepackage[dvipsnames]{xcolor}
\usepackage{mathtools}
\usepackage[normalem]{ulem}
\usepackage{soul}
\colorlet{darkred}{red!85!black}
\colorlet{darkgreen}{green!50!black}
\colorlet{darkblue}{blue!60!black}
\usepackage[
colorlinks   = true, %Colours links instead of ugly boxes
urlcolor     = darkgreen, %Colour for external hyperlinks
linkcolor    = darkblue, %Colour of internal links
citecolor   = darkred    %Colour of citations
]{hyperref}
\usepackage{braket}
\usepackage[export]{adjustbox}
\DeclarePairedDelimiter\abs{\lvert}{\rvert}%

\newcommand{\R}{\mathbb{R}}
\DeclareMathOperator{\E}{\mathbb{E}}% expected value
\def\Var{{\textrm{Var}}\,}

\newcommand{\mar}[1]{\textcolor{black}{#1}}
\newcommand{\dar}[1]{\textcolor{black}{#1}}

\begin{document}

\title{Thermalization is typical in large classical and quantum harmonic systems
}

\author{Marco Cattaneo}
\email{marco.cattaneo@helsinki.fi}
\affiliation{QTF Centre of Excellence,  
Department of Physics, University of Helsinki, P.O. Box 43, FI-00014 Helsinki, Finland}

\author{Marco Baldovin}
\affiliation{Institute for Complex Systems, CNR, 00185, Rome, Italy}

\author{Dario Lucente}
\affiliation{Department of Mathematics \& Physics, University of Campania “Luigi Vanvitelli”, 81100, Caserta, Italy}

\author{Paolo Muratore-Ginanneschi}
\affiliation{ 
Department of Mathematics and Statistics, University of Helsinki, P.O. Box 68, FI-00014 Helsinki, Finland}

\author{Angelo Vulpiani}
\affiliation{Dipartimento di Fisica, Università di Roma “La Sapienza”, P.le Aldo Moro 5, 00185, Rome, Italy}

\date{\today}

\begin{abstract}We establish an analytical criterion for dynamical thermalization within harmonic systems, applicable to both classical and quantum models. {Specifically, we prove that thermalization of various observables—such as particle energies in physically relevant random quadratic Hamiltonians—is typical for large systems ($N\gg1$) with initial conditions drawn from the microcanonical distribution.} Moreover, we show that thermalization can also arise from non-typical initial conditions, where only a finite fraction of the normal modes is excited. \mar{A different choice of initial conditions, such as all the initial energy localized in a single particle, instead leads to energy equipartition without thermalization.} \mar{Since the models we consider are integrable}, our findings provide a general dynamical basis for an approach to thermalization that bypasses chaos and ergodicity, focusing instead \mar{on the physical requirement that thermodynamic observables depend on} a large number of normal modes, and build a bridge between the classical and quantum theories of thermalization.
\end{abstract}
\maketitle

\textit{Introduction}.--%The emergence of thermalization in closed physical systems is among the most subtle phenomena in theoretical physics, stemming from the apparent contradiction between the time independence of the laws governing both classical and quantum dynamics and the irreversibility implied by thermal equilibrium. The thermalization of a gas of classical particles can be explained through statistical mechanics and kinetic theory, which were developed roughly 150 years ago \cite{huang2008statistical}. Whether or not chaos in the system dynamics and ergodicity are necessary ingredients for thermalization is, however, still a matter of debate \cite{Gaspard_1998,Bricmont2001,castiglione2008chaos,Frigg2024,Bassi2024}. The study of thermalization in isolated quantum systems is, in contrast, relatively recent \cite{}, even if some related works can be found in the first few years of the quantum theory \cite{}. Contrary to the classical case, usual arguments for thermalization in the quantum regime do not focus on the system dynamics, but rely on the Eigenstate Thermalization Hypothesis (ETH) \cite{}, which gives predictions on the behavior of the matrix elements of relevant observables evaluated in the system eigenstates. 
%
%In this Letter, we focus on an approach to thermalization that is inspired by Boltzmann's original ideas \cite{} and Khinchin's considerations \cite{Khinchin1949} (which, of course, were developed for isolated classical systems): thermalization can be explained by statistical arguments (e.g., the central limit theorem) based on the presence of thermodynamically many degrees of freedom in the expressions for the dynamical variables \footnote{``Observables'' in the language of quantum theory; we will use both ``dynamical variable'' and ``observable'' interchangeably throughout the work.} of interest. In particular, we consider quadratic models, which are a paradigmatic example of integrable and separable systems \footnote{Note that this terminology may be problematic when speaking about both classical and quantum systems, as in the quantum regime ``integrable'' may be defined in a stricter sense. Throughout the paper, an ``integrable system'' is a system of $N$ particles with $N$ conserved quantities, according to Gogolin and Eisert \cite{Gogolin2016}.}, whose thermalization cannot be explained through a chaotic dynamics. The dynamical variables we investigate are then functions of thermodynamically many normal modes of the integrable system. 
There is unanimous consensus that statistical mechanics (SM) captures in a precise way the behavior of macroscopic observables in both classical (CM) and quantum (QM) mechanics \cite{PelL2011,huang2008statistical,landau,ma1985statistical}. However, there is no complete agreement on the underlying reasons for this success.
%Among the many examples of this success  we mention the possibility to describe the probability distribution of the single-particle velocity 
%in an ideal gas, as well as the detailed behavior of phase transitions, and critical phenomena see e.g. .
%In spite of these achievements,  the consensus about the actual reasons for such a success is not complete. 
For instance, whether or not chaos in the system dynamics and ergodicity are necessary ingredients for thermalization in CM is still a matter of debate \cite{GutM1990,Gaspard_1998,Bricmont2001,castiglione2008chaos,sklar1993physics,BriJ2022,Bassi2024,Baldovin2024found}. Thermalization in QM is usually explained through the Eigenstate Thermalization Hypothesis (ETH) \cite{Deutsch1991,Srednicki1994,Rigol2008,dalessio2015,deutsch2018}, which has not been proven analytically for general interacting quantum systems, or through \textit{typicality} \cite{Tasaki1998,Gemmer2003,Goldstein2006,Popescu2006,Reimann2008,Linden2009,Reimann2013,Muller2015}, i.e., the concept that thermalization is overwhelmingly likely to occur for most configurations or states, given a sufficiently large number of degrees of freedom. 
%We cannot enter into the details of the old and delicate  debate on the foundations of the statistical mechanics.

Integrable systems \footnote{Note that the definition of ``integrable system'' is problematic in QM \cite{Gogolin2016}, while in CM we refer to systems with a maximal number of conserved quantities.} are not expected to thermalize, and their equilibration can be understood through Generalized Gibbs Ensembles (GGEs) \cite{Rigol2006,Rigol2007,Vidmar2016,Spohn2020}, which can be combined with ETH in the quantum case, see e.g.  \cite{Alba2015,Magan2016,Nandy2016,Leblond2019,Gluza2019,Lydzba2024,Tasaki2024} (see also \cite{Ptaszynski2024} for a different open system perspective). In CM we can address their thermalization, despite the lack of ergodicity, by following the lines of the seminal works of Khinchin \cite{Khinchin1949}, and then Mazur and van der Linden \cite{Mazur1963}, based on the idea that the very basic ingredients of SM are the large number of degrees of freedom and the global nature of the observables, regardless of the presence of chaos \cite{castiglione2008chaos}. This approach is further connected to the central limit theorem for the sum of trigonometric functions \cite{KacM1946}.
 The dynamical variables \footnote{``Observables'' in the language of quantum theory; we will use these terms interchangeably throughout the work.} that can display thermalization must be functions of many independent degrees of freedom, while the conserved quantities of the system are not. These concepts have {influenced some studies of SM for specific integrable systems of classical harmonic oscillators \cite{Mazur1960,Ford1965,Baldovin2023}, as well as numerical analyses for the linear chain \cite{Jin2013,Cocciaglia2022} and the Toda model \cite{Baldovin2021}.}

Inspired by these ideas, in this Letter we focus on thermalization in both classical and quantum harmonic systems of $N$ particles, which are a paradigmatic example of separable (and classically integrable) models whose motion is described by a collection of free \textit{normal modes}. Our approach is rooted in the system dynamics, as we address the time average of observables $O$ defined as
\begin{equation}
\label{eqn:timeAv}
    \overline{O}_t = \frac{1}{t}\int_0^t ds \, O(s),
\end{equation}
where in the quantum case $O(s)$ corresponds to the mean value of the observable on the system state at time $s$. If the long-time limit $\overline{O}_\infty$\mar{, for almost all initial conditions,} is equal to the ensemble average of $O$ over the microcanonical ensemble \cite{PelL2011,huang2008statistical}, which we define as $\langle O \rangle$, then we say that $O$ thermalizes (in QM the time average is not always required, see e.g. \cite{Banuls2011}).

\mar{In CM, the time variance of \textit{microscopic} observables---such as single-particle energies--- does not vanish. The time variance of \textit{macroscopic} observables, in contrast, is expected to vanish after the thermalization time \cite{landau,ma1985statistical}. In QM, also the time fluctuations of microscopic observables vanish if the system is large, provided the latter satisfies ETH \cite{dalessio2015}.}
In this work, \mar{whenever we refer to ``thermalization'' or ``equilibration'' we also check that the above prescriptions for the time variance hold.}

The observables we are interested in are the energies of some modes of the system, which are functions of many normal modes.
\mar{As we deal with harmonic systems,} we obtain the same expression for the long-time average of these energies in both CM and QM\mar{. We prove} that the conditions for their thermalization are overwhelmingly likely to occur for $N\gg1$ assuming we draw the initial conditions from the microcanonical distribution. In other words, we establish that thermalization is \textit{typical} in both classical and quantum large harmonic systems. Interestingly, we prove that thermalization still occurs even if the initial conditions are not typical, specifically when only a fraction $N^*$ of the normal modes is initially excited, provided $N^*\gg 1$. Different initial conditions, such as initial excitations uniformly distributed over the normal modes, irrespectively of their energies, also lead to energy equipartition for large $N$. However, in these cases, the equipartition value may not correspond to the one predicted by the  microcanonical ensemble average.

Our results provide a solid analytical foundation for Khinchin's approach to thermalization, grounded in the system dynamics.  To the best of our knowledge, this is the first time these ideas have been extended to the quantum regime, demonstrating their applicability to both classical and quantum systems in a comparable manner. Moreover, our findings illustrate in a clear and rigorous way how thermalization and energy equipartition are very general concepts, even in physical models where chaos does not play any role, such as harmonic systems, assuming we deal with a large number of degrees of freedom and avoid pathological initial conditions.

\textit{Model}--Any Hamiltonian of a $N$-particle harmonic system can be diagonalized in the basis of the normal modes and written as (all the summations go from 1 to $N$)
\begin{equation}
\label{eqn:DiagHamNormalModes_first}
    H = \sum_k \omega_k o_k^* o_k.
\end{equation}
$\omega_k$ are the \textit{eigenfrequencies} of the normal modes. For simplicity we assume $\omega_k>0$, and that the normal modes are non-degenerate: there is no $j\neq k$ such that $\omega_j=\omega_k$. This is a standard assumption, for instance, in the theory of quantum thermalization \cite{Gogolin2016,dalessio2015} and is usually justified by (even tiny) disorder in many-body systems. \mar{We point out that the incommensurability of the frequencies, which may be used to prove that the motion of harmonic oscillators is ergodic \cite{Giorgilli2022}, is not a requirement of our model.} $o_k$ and $o_k^*$ are symbols in a notation that is agnostic with respect to CM or QM. Specifically, in CM $o_k=z_k$, $o_k^*=z_k^*$, where $z_k=(\omega_k q_k+i p_k)/\sqrt{2\omega_k}$ and $z_k^*=(\omega_k q_k-i p_k)/\sqrt{2\omega_k}$ are the complex canonical variables of the normal modes \cite{Strocchi1966}. %, whose Poisson brackets satisfy $\{z_j,z_k^*\}=i\delta_{jk}$. 
In QM, $o_k=a_k$, $o_k^*=a_k^\dagger$,  where $a_k^\dagger$ and $a_k$ are the standard bosonic or fermionic ladder operators.  Crucially, both in CM and QM the evolution of these quantities is given by $o_k(t)=e^{-i\omega_k t}o_k(0)$, $o_k^*(t)=e^{i\omega_k t}o_k^*(0)$.

Harmonic systems are \textit{separable} \cite{GutM1990}, i.e., $H$ can be divided into the sum of $N$ constants of motion: $H=\sum_k E_k$. The full state of this system will never thermalize, as energy cannot be exchanged between the normal modes. Our goal, however, is to focus on dynamical variables that are a function of many normal modes. 

\begin{figure*}
    \centering
    \includegraphics[scale=0.36]{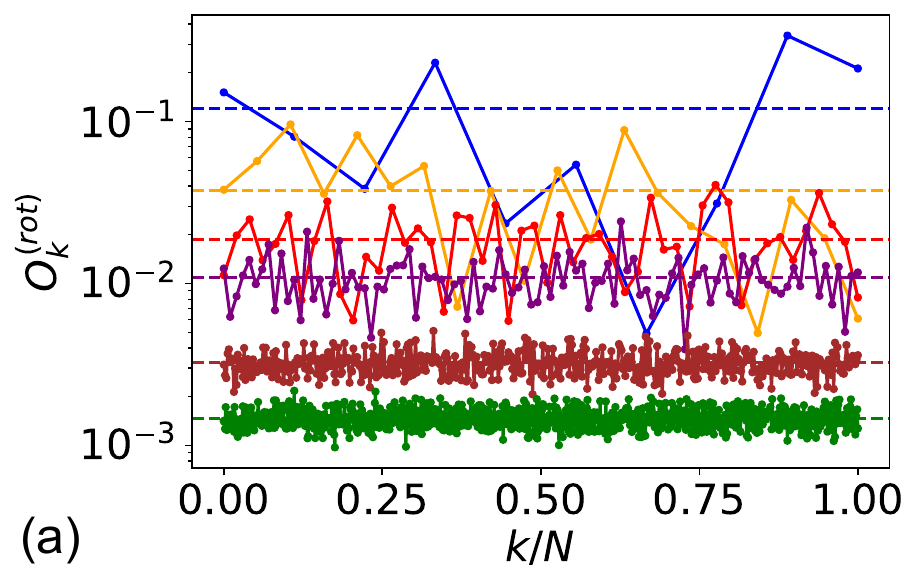}
    \includegraphics[scale=0.36]{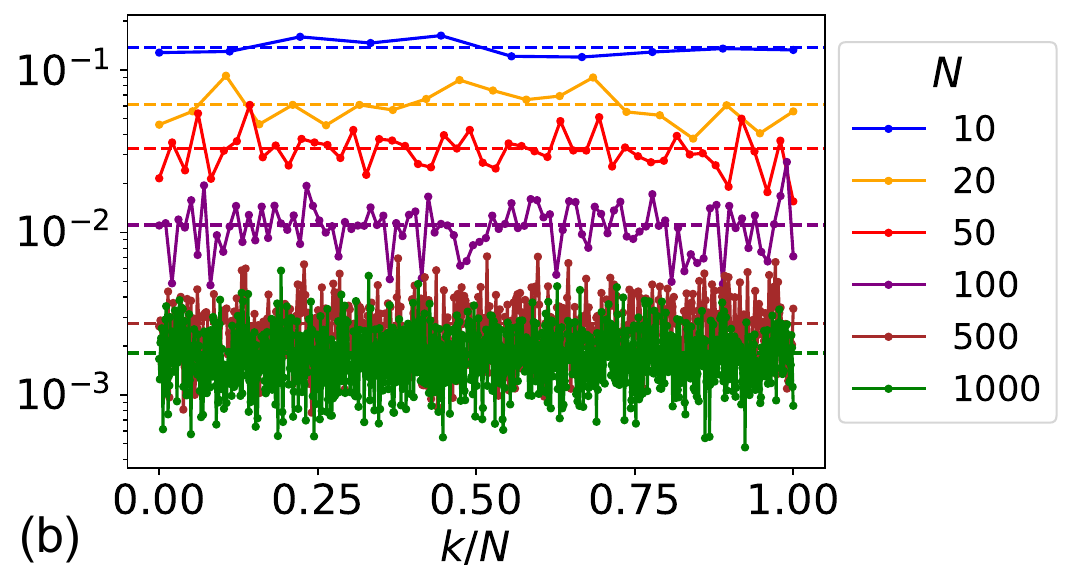}
    \includegraphics[scale=0.36]{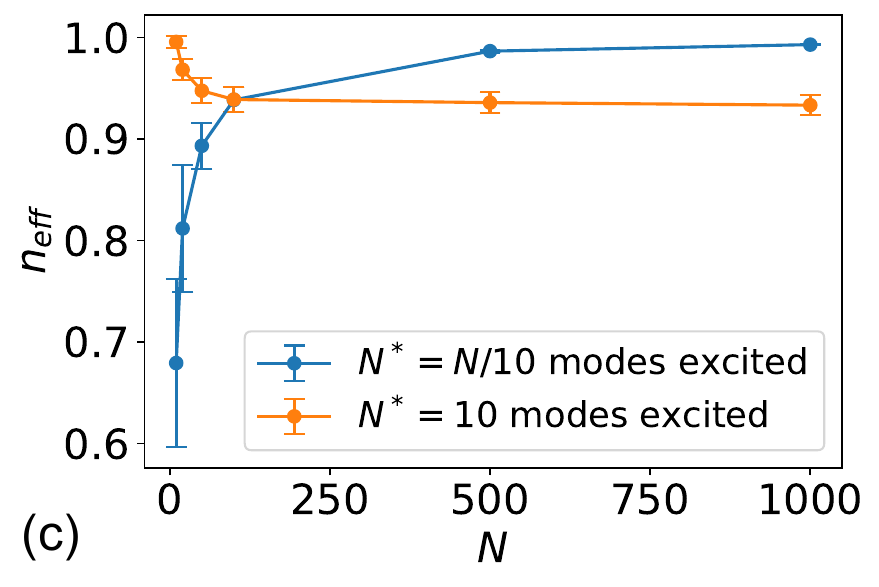}
        \caption{Numerical results for the energy of randomly rotated modes, with random eigenfrequencies $\omega_j$  sampled uniformly over $[1,10]$.  (a): Distribution of $O^{(rot)}_k$ over $k$ for a single realization of $C$ (Eq.~\eqref{eqn:Energy_Rot_modes}). The initial conditions are chosen in such a way that the first $N^*$ modes have the same energy, while the others are not excited. This is equivalent to the restricted \dar{classical} microcanonical distribution discussed in~\cite{Supplemental}. The dashed lines are the theoretical predictions given by Eq.~\eqref{eqn:meanVMicroAvRot}. (b): same as (a), but only 10 normal modes are excited in the initial state independently of $N$. (c): $n_\text{eff}$ (Eq.~\eqref{eqn:neff})  as a function of $N$ for the two kinds of initial conditions depicted in (a) and (b) averaged over 50 different realizations of $C$ (error bars = 1 std. dev.). \dar{Data points are joined by a line for visual guidance only.}}
    \label{fig:randomRot}
\end{figure*}

\textit{Randomly rotated modes}.--We first consider a rotation of the normal modes:
\begin{equation}
    \label{eqn:randomRot}
    \tilde{o}_j = \sum_k C_{jk} o_k,\qquad \tilde{o}^*_j = \sum_k C^*_{jk} o^*_k.
\end{equation}
$C_{jk}$ is a generic orthogonal or unitary matrix. The Hamiltonian written in the basis of rotated modes is not separable anymore:
\begin{equation}
\label{eqn:HamRandomRotRotModes}
    H = \sum_k \tilde{E}^{(rot)}_k + \sum_{k,k':k\neq k'} \mu_{kk'}\tilde{o}_k^* \tilde{o}_{k'}, 
\end{equation}
with $\tilde{E}^{(rot)}_k = \mu_{kk} \tilde{o}_k^* \tilde{o}_{k}$ and $\mu_{kk'}=\sum_{j}C_{kj} C_{k'j}^*\omega_j$. Intuitively, energy can be exchanged between the rotated modes, so we are interested in studying the thermalization of the energies $\tilde{E}^{(rot)}_k$ in the regime $N\gg 1$. We consider their long-time average and obtain
\begin{equation}
    \label{eqn:Energy_Rot_modes}
    O^{(rot)}_k:=\overline{\tilde{E}^{(rot)}_k}_\infty = \sum_j \abs{C_{kj}}^2 p_{0,j} \sum_{j'} \abs{C_{kj'}}^2  \omega_{j'},
\end{equation}
where the initial \textit{populations} are defined as $p_{0,j}=\abs{z_j(0)}^2$ for CM and $p_{0,j}=\Tr[a_j^\dagger a_j\rho_0]$ for QM.
For the derivations of all the results in the main text we refer the readers to the Supplemental Material %(SM) 
\cite{Supplemental}. \mar{Next, we analyze three different scenarios for the initial conditions of the normal modes.}%Note that Eq.~\eqref{eqn:Energy_Rot_modes} depends on $k$ only through $\abs{C_{kj}}^2$, so if $\abs{C_{kj}}^2\approx 1/N$ for all $j,k$ energy is equipartitioned. 

%Moreover, it can be easily shown that if we draw the initial populations $p_{0,j}$ according to the microcanonical distribution (both in CM and QM), we obtain:

\textit{Case 1: initial conditions from the microcanonical distribution.}--If we now assume that (i) the initial populations are extracted according to the microcanonical distribution (both in CM and QM) and (ii) the rotation matrix $C_{jk}$ is drawn uniformly over the sphere, it can be shown that thermalization %to the equipartitioned state above 
is typical. First one easily notices that~\cite{Supplemental}
\begin{equation}
    \label{eqn:meanVrotMicro}
    \langle O^{(rot)}_k\rangle = \langle \tilde{E}^{(rot)}_k\rangle\,,
\end{equation}
where $\langle \cdot \rangle$ denotes a microcanonical average. {On the left-hand side, this average is taken over the initial conditions, while on the right-hand side, it refers to the standard ensemble average of $\tilde{E}_k^{(rot)}$}.
%which means, the microcanonical average of the rotated energies is equal to their long-time average if we perform an additional microcanonical average over the initial conditions.
This result alone, however, is not enough:
%to claim that thermalization is typical. 
we also have to check that the fluctuations around the microcanonical average over the initial conditions vanish for large $N$, i.e., that thermalization is reached for almost any initial set of populations drawn from the microcanonical ensemble. \mar{To understand why this check is necessary,  notice that the energies of the normal modes are, trivially, thermalized on average when the initial conditions are drawn from the microcanonical distribution. However, fluctuations around this average do not vanish for large $N$, so the fraction of initial conditions leading to non-thermalized normal modes remains finite, and thermalization is therefore non-typical.}

The fluctuations are captured by
\begin{equation}
\label{eq:typicality}
\epsilon_k := \frac{ \langle (O^{(rot)}_k)^2\rangle-\langle \tilde{E}^{(rot)}_k\rangle^2}{\langle \tilde{E}^{(rot)}_k\rangle^2}.
\end{equation}
%Let us now assume that the  rotation $C_{kj}$ is drawn  uniformly over the sphere. 
Within our assumptions, the rows of $C$ can be thought of as the vectors of an orthonormal basis in $\R^N$ \footnote{For simplicity we provide all the results for a real $C$; generalizations to a complex unitary are trivial.}, which for large $N$ are uniformly distributed over the sphere \cite{Goldstein2017a}. \mar{Using their probability distribution \cite{Supplemental}, we can compute the moments of the random elements $\abs{C_{kj}}^2$. We obtain}
\begin{equation}    \label{eqn:typicalityRot_mean} 
    \E[\epsilon_k] \rightarrow 0 \text{ for }N\gg1,
\end{equation}
and
\begin{equation}    \label{eqn:typicalityRot} 
    \frac{\Var[\epsilon_k]}{(\E[\epsilon_k])^2}\rightarrow 0 \text{ for }N\gg1,
\end{equation} where $\E[\cdot]$ and $\Var[\cdot]$ refer respectively to the mean value and variance over ``disorder'', i.e., different realizations of the random rotation matrix.
In other words, for \mar{almost any set of modes obtained through the random rotation} $C$ the fluctuations $\epsilon_k$ of the {quantity} $O_k^{(rot)}$ vanish in the large-$N$ limit, hence the result.

%Eq.~\eqref{eqn:typicalityRot} can be proven under reasonable assumptions about the distribution of the eigenvalues $\omega_k$ \cite{Supplemental}.  

Moreover,
\begin{equation}
\label{eqn:meanVMicroAvRot}
    \E[\langle \tilde{E}^{(rot)}_k\rangle]= \omega_\text{av} \langle p_0\rangle_\text{av},
\end{equation} 
with $\langle p_0\rangle_\text{av}=\frac{E_\text{tot}}{N}(\omega^{-1})_\text{av}$ for CM and $\langle p_0\rangle_\text{av}=\sum_j 1/(N(1\pm e^{-\beta\omega_j}))$ for respectively fermions and bosons in QM \mar{($\beta$ is the inverse temperature in the microcanonical ensemble  \cite{Supplemental})}. We have also introduced the total energy $E_\text{tot}=\sum_j \omega_jp_{0,j}$, $\omega_\text{av}=\sum_j \omega_j/N$, and $(\omega^{-1})_\text{av}=\sum_j 1/(N\omega_j)$. Eq.~\eqref{eqn:meanVMicroAvRot} is independent of $k$, i.e, the energy of the rotated modes is equipartitioned.  However, note that the energy stored in the rotated modes at infinite time may not be equal to $E_\text{tot}$, as some may be trapped in the interaction part of Eq.~\eqref{eqn:HamRandomRotRotModes}. \mar{Note that, while QSM does not predict energy equipartition of the normal modes at low temperatures \cite{PelL2011}, Eq.~\eqref{eqn:meanVMicroAvRot} predicts equipartition for the rotated modes also in QM, as $\mu_{kk}$ in Eq.~\eqref{eqn:HamRandomRotRotModes} does not depend on $k$ for $N\gg 1$.}

\textit{Case 2: only a fraction of the normal modes is initally excited.}--Interestingly, it can be shown that Eqs.~\eqref{eqn:meanVrotMicro},~\eqref{eqn:typicalityRot} and ~\eqref{eqn:meanVMicroAvRot} hold true even if we consider microcanonical averages over a finite subset $1\ll N^*\ll N$ of the normal modes, showing that the initial conditions do not have to be typical on the full space of configurations \cite{Supplemental}. We validate the latter result through numerical simulations 
shown in Fig.~\ref{fig:randomRot}\dar{, which refer to classical systems}. We consider rotations through some complex random unitary $C$. We plot the final distribution of $O^{(rot)}_k$ over $k$ for different $N$ in Fig.~\ref{fig:randomRot}(a) and \ref{fig:randomRot}(b) and compare them with {Eq.~\eqref{eqn:meanVMicroAvRot}}, respectively when we initially excite a fraction $N^*=N/10$ of normal modes with populations {given by the (restricted) \dar{classical} microcanonical average}, and for a scenario in which only the first 10 normal modes are excited independently of $N$. In (a) we observe that the fluctuations around the mean value vanish for large $N$ and {equipartition} emerges. In (b) we are violating the condition $N^*\gg 1$, so the fluctuations do not vanish. This behavior can be captured by the normalized \textit{effective number of degrees of freedom} \cite{Baldovin2021}:
\begin{equation}
    \label{eqn:neff}
    n_\text{eff}=\frac{\exp\left(-\sum_k u_k \log u_k \right)}{N},
\end{equation}
with $u_k=O^{(rot)}_k/\sum_j O^{(rot)}_j$. The energy is equipartitioned if $n_\text{eff}= 1$. %, while $n_\text{eff}\rightarrow0$ corresponds to all the energy localized in a single mode. 
Fig.~\ref{fig:randomRot}(c) shows that $n_\text{eff}\rightarrow 1$ for $N\gg 1$ only if a fraction $N^*\gg 1$ of the normal modes is initially excited.%, as we were expecting. 

\textit{Case 3: roughly equal initial populations.}--Let us now discuss a case where the populations are not drawn from the microcanonical ensemble, but satisfy a different notion of ``typicality''. In particular, we assume that all populations are initially excited \mar{with similar magnitude}. In other words, if we introduce the average population ${p}_\text{av}=\sum_j p_{0,j}/N$, we assume $p_{0,j}/p_\text{av}=\mathcal{O}(1)$ for all $j$. This is a common initial condition describing, for instance, a situation in which only one single rotated mode is initially excited. \mar{Indeed, if we initially excite only the $\alpha$th rotated mode with some energy $E_0$, $\tilde{z}_j^*(0)\tilde{z}_{j'}(0)=E_0 \delta_{j\alpha}\delta_{j'\alpha}$ (and analogously for QM),  from Eq.~\eqref{eqn:randomRot} we obtain $\abs{z_k(0)}^2 = E_0 \abs{C_{k\alpha}}^2$, which does not depend on $k$ for large $N$.}

Moreover, let us also assume that the magnitude of the eigenfrequencies is comparable, i.e., $\omega_j/\omega_\text{av}=\mathcal{O}(1)$ for all $j$. Then, \mar{it can be shown that for $N\gg 1$ the energies of the rotated modes are typically equipartitioned for both CM and QM. However, the equipartition value does not correspond to the microcanonical one, unless all the eigenfrequencies become identical for large $N$.} 

\mar{In other words, when the initial populations of the normal modes are comparable, we typically observe equilibration towards energy equipartition without thermalization. In this scenario, thermalization cannot occur unless the system dynamics exhibit non-integrability. This is because these initial conditions constitute a measure-zero (although physically relevant) subset in the microcanonical ensemble for $N\gg1$. %Nevertheless, this subset may capture physically relevant situations, such as when all the energy is initially concentrated in a single rotated mode.
This scenario describes initial states where the observables we are studying deviate more significantly from equilibrium compared to those sampled from the microcanonical distribution. Further discussions and details on this topic can be found in Appendix A (see also \cite{Supplemental}).}

\mar{Finally, we have verified that the time fluctuations of $\tilde{E}_k^{(rot)}$ do not vanish in CM, while they do disappear for fermions and hard-core bosons in QM, provided $N\gg 1$ and for all sets of initial conditions discussed above. Moreover, time fluctuations correctly vanish also for classical systems if we consider a macroscopic observable, e.g. the sum of several $\tilde{E}_k^{(rot)}$. More details can be found in Appendix B.}

\textit{Application to \dar{Hamiltonians with delocalized eigenvectors}}--Let us now analyze a different set of physically motivated observables. We consider the Hamiltonian
\begin{equation}
\label{eqn:randomHam}
    H = \sum_{j,k=1} M_{jk} \tilde{o}_j^* \tilde{o}_k,
\end{equation}
with $M_{jk}=M_{kj}^*$. $M_{kk}$ is the self-energy of the ``local mode'' $\tilde{o}_k$ of the $k$th particle, while $M_{jk}$ is the inter-particle interaction. Eq.~\eqref{eqn:randomHam} can describe many different physical systems of interest, such as a collection of interacting harmonic oscillators, or an ensemble of superconducting qubits \cite{Blais2020} %\footnote{The qubits can be obtained through the standard Jordan-Wigner transformations on the fermionic operators, such that $\tilde{o}_j^*=\sigma_j^+$ and $\tilde{o}_j=\sigma_j^-$ \cite{coleman2015introduction}.}  
coupled in a structured way (e.g., in the topology of a quantum computer \cite{Kim2023})\mar{, or interacting with a spin bath \cite{Pekola2023}}. $H$ can be  diagonalized through a rotation $C$, which is the matrix of the normalized eigenvectors of $M$, and written in the basis of the normal modes as in Eq.~\eqref{eqn:DiagHamNormalModes_first}, whose relation with the local modes is still given by Eq.~\eqref{eqn:randomRot}. We again assume that the eigenfrequencies of the normal modes are non-degenerate.

We are interested in the long-time average of the energy of each local particle $\tilde{E}^{(loc)}_k=M_{kk} \tilde{o}_k^*\tilde{o}_k$, as this quantity may be easily accessible in an experiment  \cite{Supplemental}:
\begin{equation}
    \label{eqn:Energy_Loc_modes}
    O^{(loc)}_k:=\overline{\tilde{E}^{(loc)}_k}_\infty =  M_{kk} \sum_j \abs{C_{kj}}^2 p_{0,j}.
\end{equation}
%If $M$ is a random matrix, then $O^{(loc)}_k$ is a random variable as it depends on $C$. 
$p_{0,j}$ are the initial conditions, as previously defined. %

 Let us now assume that the eigenvectors of the matrix $M$ get \textit{delocalized} for large $N$ \cite{ORourke2016}, that is, we can effectively think of them as uniformly distributed over the unit sphere up to fluctuations vanishing with $N$. Then, the rows of $C$ are distributed \mar{uniformly over the sphere} and we can calculate the moments of $\abs{C_{kj}}^2$ \cite{Supplemental}. Different classes of physically relevant random matrices have been proven to have delocalized  eigenvectors for large $N$, such as the GOE and GUE ensembles \cite{Anderson2009,ORourke2016}, the Rosenzweig-Porter model \cite{Altland1997,Venturelli2023}, symmetric and Hermitian matrices \cite{Erdos2009,Erdos2009a}, and more \cite{Rudelson2015,Benigni2022}. So, it is conjectured that eigenvector delocalization is a typical property of random matrices for large $N$ \cite{Rudelson2017}. %Moreover, almost any orthonormal basis of eigenvectors is delocalized for $N\rightarrow\infty$ \cite{Goldstein2017a}. %A hands-on way to study whether the eigenvectors of $M$ get delocalize for $N\gg 1$ or not is to check that their Inverse Participation Rate (IPR) \cite{} renormalized by $N$ is independent of $N$. The IPR is a widely used tool to study thermalization and chaos \cite{}.

Then, using $M_{kk}=\sum_j \abs{C_{kj}}^2\omega_j$, $O^{(loc)}_k$ can be easily transformed into the expression for $O^{(rot)}_k$ in Eq.~\eqref{eqn:Energy_Rot_modes} and it can be shown that the thermalization of $O^{(loc)}_k$ is also typical for initial conditions drawn uniformly from the microcanonical distribution: Eq.~\eqref{eqn:typicalityRot} still holds \cite{Supplemental}. \mar{Similar results can be found also in the case of a single initially excited local mode, assuming all the self-energies are equal for large $N$. We refer the interested readers to the discussion in Appendix C.}

Finally, we remark that a similar \mar{reasoning applies to} the analogous case of quadratic Hamiltonians of the form $H=\sum_j p_j^2/2 + \sum_{jk} v_{jk} q_jq_k$, which is of great relevance for CM \cite{Mazur1960,Ford1965}. For further details we refer to %the SM 
\cite{Supplemental}, where we also prove \mar{that the eigenvectors of the linear chain \cite{Cocciaglia2022,Baldovin2023} get delocalized for $N\gg 1$. \dar{This holds independently of the boundary conditions of the chain, and it is true for both classical and quantum modes.}} \dar{At the same time, our results do not hold anymore for Hamiltonians with localized eigenvectors, such as lattices with disorder inducing Anderson-like localization.}

\textit{Conclusions}--
In this Letter, we have shown that thermalization and energy equipartition are overwhelmingly likely to occur for the energies of almost any set of modes in both classical and quantum harmonic systems with a sufficiently large number of degrees of freedom and non-pathological initial conditions. These modes include those derived from random rotations of the normal modes and local modes of quadratic random Hamiltonians, which \mar{describe several paradigmatic models in} both classical and quantum mechanics. We have considered two different notions of ``typical" initial conditions. The first involves ``populations" of the normal modes (or of a subset thereof) drawn from the microcanonical distribution, representing a scenario where the total energy of the system is fixed before being distributed among the normal modes. The second notion involves populations drawn from a peaked distribution, representing, for example, a scenario where only a single rotated mode is initially excited. %, as in the paradigmatic Fermi-Pasta-Ulam-Tsingou (FPUT) experiment \cite{Berman2005,Gallavotti2008}. 
In the first scenario, thermalization and energy equipartition are typical \dar{(see Eqs.~\eqref{eqn:meanVrotMicro},~\eqref{eqn:typicalityRot} and~\eqref{eqn:meanVMicroAvRot})}. In the second case, energy equipartition is typical at a value that may differ from that predicted by the microcanonical average \dar{(see Appendix A)}. 

Our results, supported by various numerical simulations, offer robust analytical evidence for a conceptualization of thermalization based solely on the large number of degrees of freedom and typical initial conditions. Additionally, we have shown for the first time that this concept holds true for both classical and quantum mechanics fundamentally due to  the same reasons. \mar{Our findings may thus provide an alternative way to understand thermalization, with significant implications, for instance,  for the quantum-to-classical transition in statistical mechanics. For example, one may view the general concepts and ideas behind ETH as exhibiting a similar spirit to Khinchin's approach to thermalization \cite{pappalardiLectureNotes}, which we have consolidated and demonstrated for a broad class of classical and quantum models. Then, a natural and promising direction for future work would be to investigate the connections between our framework and various results on the emergence of ETH at different levels and with different implications in integrable systems \cite{Rigol2011,Magan2016,Lydzba2020,Lydzba2023,Lydzba2024,Tasaki2024}. This may provide a more unified understanding of thermalization across the classical and quantum domains, which have traditionally been treated through very distinct approaches. Moreover, it would be interesting to connect our results with other recent works that address thermalization in integrable systems relevant to current quantum technologies \cite{Pekola2023,Ptaszynski2024,Pekola2024}.}

\textit{Acknowledgments}--
The authors would like to thank Erik Aurell, Cecilia Chiaracane, John Goold, Giacomo Gradenigo, Bayan Karimi, Antti Kupiainen, Patrycja {\L}yd{\.z}ba, Silvia Pappalardi, Jukka Pekola, Luca Peliti, and Massimiliano Viale for interesting discussions related to this work. MC acknowledges funding from the Research Council of Finland through the Centre of Excellence program grant 336810 and the Finnish Quantum Flagship project 358878 (UH), and from COQUSY project PID2022-140506NB-C21 funded by MCIN/AEI/10.13039/501100011033.
MB was supported by ERC Advanced Grant RG.BIO (Contract No. 785932).

\appendix

\subsection{Appendix A: Initial conditions with roughly equal populations}

\mar{Given ${p}_\text{av}=\sum_j p_{0,j}/N$, we assume $p_{0,j}/p_\text{av}=\mathcal{O}(1)$ for all $j$. Moreover, $\omega_j/\omega_\text{av}=\mathcal{O}(1)$ for all $j$. Then,}
for $N\gg1$ we obtain \cite{Supplemental}
\begin{equation}
    \label{eqn:meanVrandomRotNonMicro}
\E[O^{(rot)}_k] = \omega_\text{av}p_\text{av} \left(1+ \mathcal{O}\left(1/N\right)\right),
\end{equation}
The above expression does not depend on $k$, i.e., the energy is on average equipartitioned among the rotated modes. Furthermore, we find $\Var[O^{(rot)}_k] = \mathcal{O}\left(\omega_\text{av}^2p_\text{av}^2/N\right)
$,
which immediately implies
\begin{equation}
\label{eqn:finalScaling_rot}
    \frac{\Var[O^{(rot)}_k]}{(\mathbb{E}[O^{(rot)}_k])^2}=\mathcal{O}\left(\frac{1}{N}\right)\rightarrow 0 \text{ for }N\rightarrow\infty, 
\end{equation}
proving that energy equipartition is typical. It is also possible to show that
\begin{equation}
\label{eqn:finalScaling_rot_mean}
\E[O^{(rot)}_k] \rightarrow \frac{E_\text{tot}}{N} \text{ for }N\rightarrow\infty,
\end{equation}
for typical initial conditions~\cite{Supplemental}. \mar{We have introduced the total energy in the system $E_\text{tot}=\sum_j \omega_jp_{0,j}$.} 

\mar{We now compare Eq.~\eqref{eqn:meanVrandomRotNonMicro} with the corresponding value obtained for initial conditions drawn from the classical microcanonical distribution, Eq.~\eqref{eqn:meanVMicroAvRot}. Using Eq.~\eqref{eqn:finalScaling_rot_mean}, for CM their difference can be written as
\begin{equation}
\frac{E_\text{tot}}{N} \left(\omega_\text{av}
    (\omega^{-1})_\text{av}-1\right).  
\end{equation} 
Therefore, in CM the two values coincide only if} $\omega_\text{av}\approx1/(\omega^{-1})_\text{av}$, i.e., the distribution of the eigenfrequencies must be peaked, with relative fluctuations vanishing for large $N$.

\mar{
The key difference between the microcanonical distribution and the ``roughly equal'' initial conditions $p_{0,j}$ lies in the weighting factor $1/\omega_j$. Suppose the eigenfrequencies are uniformly distributed over the interval $[1,2]$. Setting $p_{0,j}=p_0$ independently of $j$  neglects the factor $1/\omega_j$, which is nonetheless within the standard microcanonical fluctuations of the $j$th mode population \cite{Supplemental}. However, this  choice of $p_{0,j}$ for all the modes systematically biases the initial conditions relative to the microcanonical distribution: modes with higher frequencies $(\omega_j\geq 1.5)$ will, on average, be more populated than in the microcanonical ensemble, while lower-frequency modes will be underpopulated. This is why this choice represents a zero-measure subset of the microcanonical ensemble in the limit $N\gg 1$.
}

%Eq.~\eqref{eqn:meanVrandomRotNonMicro} (roughly {equal} initial populations) the equipartition value is equal to the microcanonical average in Eq.~\eqref{eqn:meanVMicroAvRot} if and only if $\langle p_0\rangle_\text{av}=p_\text{av}$. For instance, for CM this condition can be transformed into $\omega_\text{av}\approx 1/(\omega^{-1})_\text{av}$, i.e., the distribution of the eigenfrequencies must be peaked, with relative fluctuations vanishing for large $N$.

\subsection*{Appendix B: Time fluctuations}

\begin{figure}
    \centering
    \includegraphics[scale=0.45]{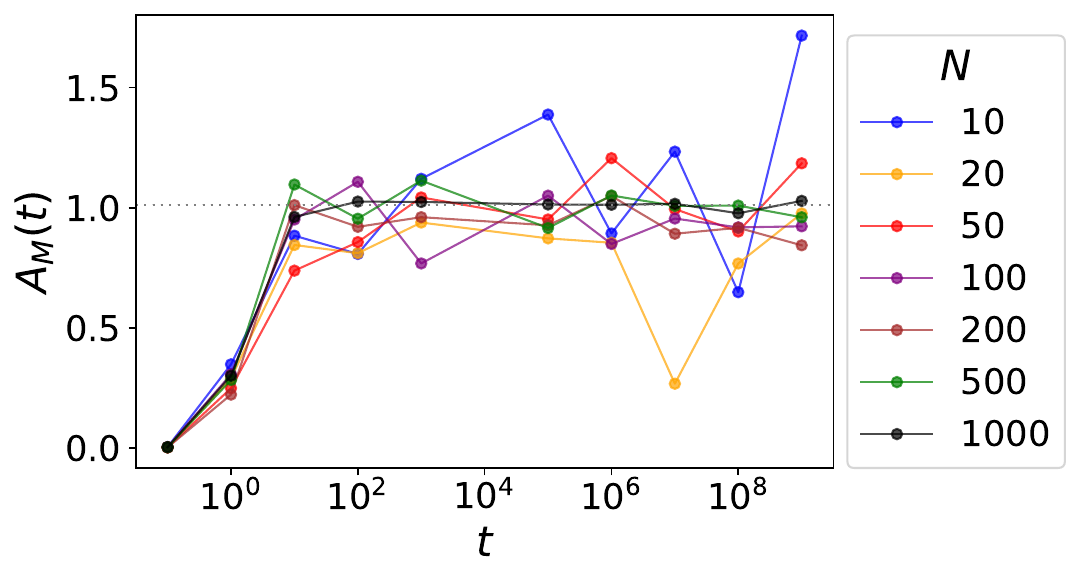}
    \caption{\mar{$A_M$ from Eq.~\eqref{eqn:macroObs} as a function of time, with $\mathsf{K}=(N/2,\ldots,N)$., for different values of $N$. The initial energy is concentrated in the first rotated mode. The eigenfrequencies are sampled uniformly from $[1,3]$. The grey dotted line depicts the theoretical prediction for the equilibration value derived from Eq.~\eqref{eqn:finalScaling_rot_mean}.}}
    \label{fig:magn_FPUT}
\end{figure}

\begin{figure*}
    \centering
    \includegraphics[scale=0.36]{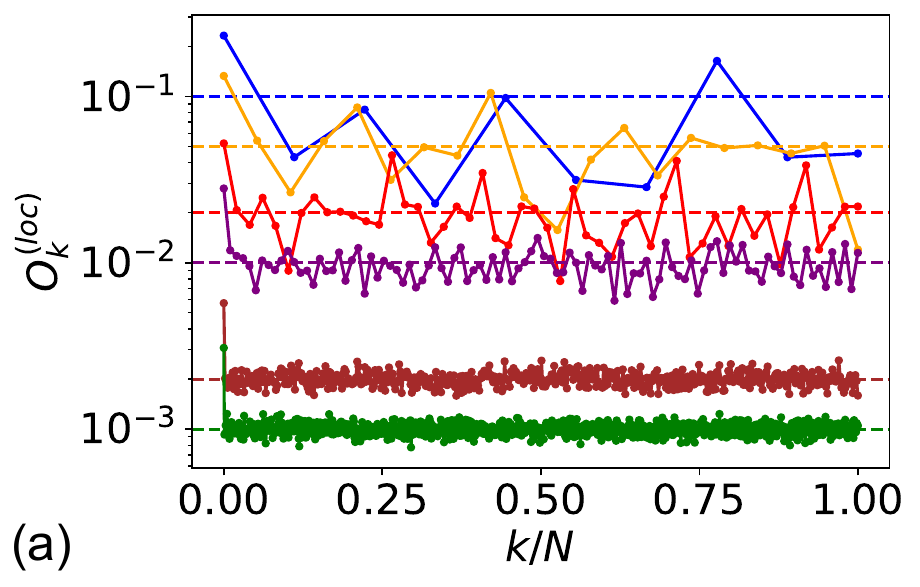}
    \includegraphics[scale=0.36]{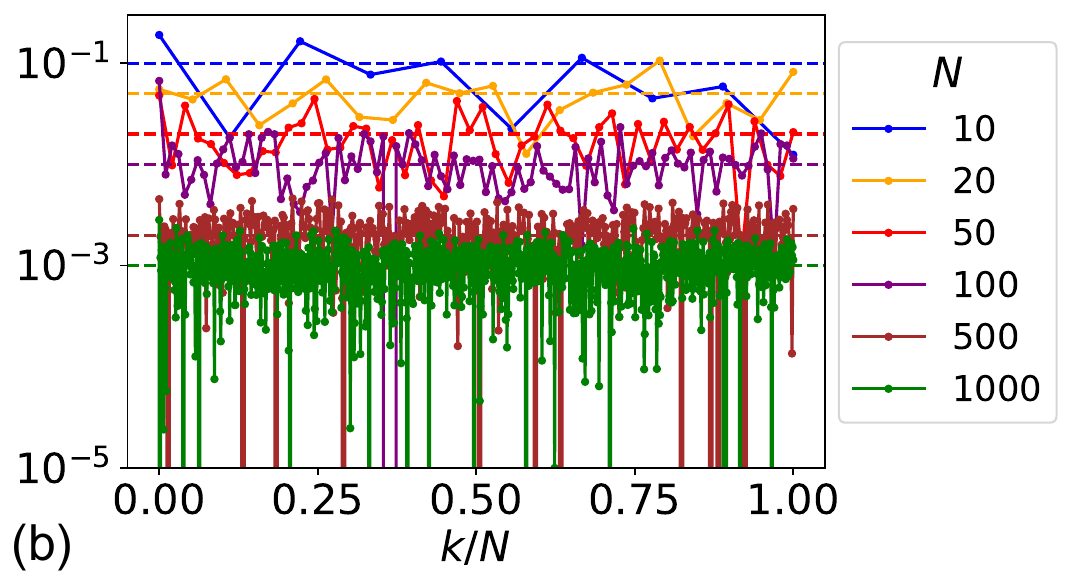}
    \includegraphics[scale=0.36]{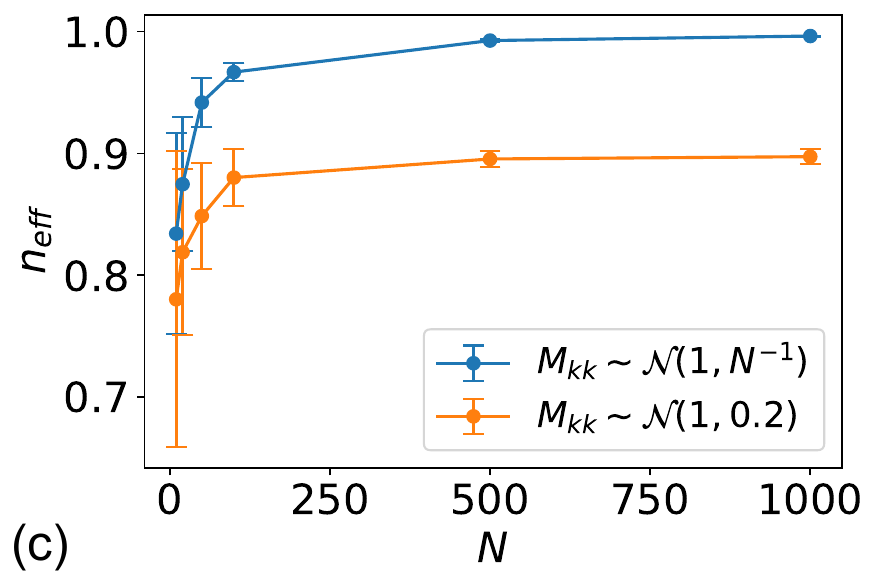}
    \caption{Numerical results for the energy of the local particles of a random $H$ in Eq.~\eqref{eqn:randomHam}. We initially excite a single local mode ($k=1$) in a Fermi-Pasta-Ulam-Tsingou (FPUT) fashion \cite{Berman2005,Gallavotti2008}, so that the $p_{0,j}$ are (roughly) randomly distributed among the normal modes. (a): Distribution of $O^{(loc)}_k$  over $k$  for a single realization of $M$ and with $M_{jk}\sim\mathcal{N}(0,0.1)$, $M_{kk}\sim\mathcal{N}(1,N^{-1})$. (b): same as (a) but with $M_{kk}\sim\mathcal{N}(1,0.2)$. (c): $n_\text{eff}$ (Eq.~\eqref{eqn:neff}) as a function of $N$ in the two scenarios depicted in (a) and (b) averaged over 50 different realizations of $M$ (error bars = 1 std. dev.). \dar{Data points are joined by a line for visual guidance only.}}
    \label{fig:randomHam}
\end{figure*}
\subsubsection*{Microscopic observables}
\mar{In this section we study the time variance of the energies of each rotated mode. We make the additional assumption that there are no gap degeneracies in the spectrum of the eigenfrequencies, which is usual in the context of ETH \cite{dalessio2015,Lydzba2023}.
Using the definition in Eq.~\eqref{eqn:HamRandomRotRotModes}, for CM we observe \cite{Supplemental}
\begin{equation}
    \overline{(\tilde{E}^{(rot)}_k)^2}_\infty \rightarrow 2 \left(\overline{\tilde{E}^{(rot)}_k}_\infty\right)^2\text{ for }N\rightarrow\infty.
\end{equation}
Therefore, the time fluctuations do not vanish in the classical regime. 
}

\mar{In contrast, for fermions and hard-core bosons in QM we obtain \cite{Supplemental}
\begin{equation}
    \overline{(\tilde{E}^{(rot)}_k)^2}_\infty \leq \left(\overline{\tilde{E}^{(rot)}_k}_\infty\right)^2(1 + \mathcal{O}(1/N)),
\end{equation}
for large $N$. As a consequence, the time fluctuations vanish in the quantum case for $N\gg 1$.
}

\subsubsection*{Macroscopic observables}
\mar{As a macroscopic observable, we consider any quantity that can be written as a sum of single-particle energies of the rotated modes, or single-particle occupations number of the rotated modes \cite{Goldstein2010}. Let us introduce the observable 
\begin{equation}
\label{eqn:macroObs}
    A_M = \sum_{k\in \mathsf{K}} \tilde{E}^{(rot)}_k,
\end{equation}
where $\mathsf{K}$ is a set containing  $N^*$ rotated modes, which for large $N$ is a finite fraction of all the modes: $N^*\gg1$ if $N\gg 1$. Then, for both CM and QM it can be shown that \cite{Supplemental}
\begin{equation}\label{eqn:fluctuationsAM}
    \overline{A_M ^2}_\infty= \overline{A_M}_\infty^2 (1+\mathcal{O}(1/N^*))\text{ for }N^*\gg 1. 
\end{equation}
Time fluctuations of macroscopic observables thus correctly vanish in the thermodynamic limit. Note we would obtain the same results if we replaced  $\tilde{E}^{(rot)}_k$ with the occupation number $\tilde{o}_k^* \tilde{o}_{k}$ in Eq.~\eqref{eqn:macroObs}.
}

\mar{To test this prediction numerically, we study $A_M$ as the sum of the energies from mode $k=N/2$ to $k=N$ ($N^*=N/2$). We consider a scenario where only the first rotated mode is initially excited, thus $A_M(0)=0$, and study $A_M(t)$ for different $N$. Its time evolution is plotted in Fig.~\ref{fig:magn_FPUT}. $A_M(t)$ grows in time until approaching the equilibration value, and time fluctuations are clearly suppressed for larger $N$. If, instead, we drew the initial conditions from the microcanonical ensemble, $A_M(0)$ would be closer to the thermalization value, and fluctuations around it would vanish for $N\gg1$.} 

\mar{This observation is consistent with Khinchin's original idea in classical SM, where thermalization arises solely from the choice of observable and the structure of the microcanonical ensemble, independently of time evolution. If an observable meets the requirements for thermalization, it will thermalize for most initial conditions (with probability approaching 1 in the microcanonical ensemble), at any time. The notion of typicality in QM is also based on similar ideas \cite{Baldovin2024found}.  For technical details on the connection between Khinchin's results and our findings, see \cite{Supplemental}.}

\subsection*{Appendix C: Energy equipartition for a random quadratic Hamiltonian}

\mar{In this section, we study the long-time average of the observables $\tilde{E}_k^{(loc)}$ introduced in Eq.~\eqref{eqn:Energy_Loc_modes}, for a random quadratic Hamiltonian given by Eq.~\eqref{eqn:randomHam}.} Note that, if $M_{kk}=M_0$ for all $k$ (``identical particles'', e.g., a collection of resonators with equal frequency \cite{Blais2020}) and, crucially, $\abs{C_{jk}}^2\approx 1/N$, energy is equipartitioned among the particles.

\mar{Instead of drawing the initial conditions from the microcanonical ensemble, we now assume} a peaked distribution of the populations, i.e.,  $p_{0,j}/p_\text{av}=\mathcal{O}(1)$ with $p_\text{av}=\sum_j p_{0,j}/N$. 
We obtain \cite{Supplemental}
\begin{equation}
\label{eqn:meanVlocal}
     \mathbb{E}[O^{(loc)}_k] = \mathbb{E}[M_{kk}]\,p_\text{av},
\end{equation}
\begin{equation}
\label{eqn:varLocMod}
\begin{split}
    \Var[O^{(loc)}_k] =&p_\text{av}^2\left( \Var[M_{kk}]+\mathcal{O}\left(\frac{\Var[M_{kk}]}{N}\right)\right.\\
    &\left.+\mathcal{O}\left(\frac{(\mathbb{E}[M_{kk}])^2}{N}\right)\right).
\end{split}
\end{equation}
Next, we assume $\mathbb{E}[M_{kk}]=M_0$ for all $k$ and the fluctuations around the mean value vanish for large $N$, i.e., $\Var[M_{kk}]=1/N^{\alpha}$ with $\alpha>0$. This means that the particles must be identical for $N\rightarrow\infty$. \mar{Note that, since we assume that all the $\omega_j$ are of the same order, this is consistent with delocalized eigenvectors for large $N$, as $M_{kk}=\sum_{j}\abs{C_{kj}}^2\omega_j$.} Then,
\begin{equation}
\label{eqn:fluctuationsLocalMod}
    \frac{\Var[O_\infty^{(k)}]}{(\mathbb{E}[O_\infty^{(k)}])^2}=\mathcal{O}\left(\frac{1}{N^{\alpha}}\right)\rightarrow 0 \text{ for }N\rightarrow\infty,
\end{equation}
that is, energy is equipartitioned among the particles.

We numerically illustrate the emergence of equipartition in the latter scenario  by first generating a random matrix with elements drawn according to $M_{kk}\sim\mathcal{N}(1,N^{-1})$ and $M_{jk}\sim\mathcal{N}(0,0.1)$ %, describing a collection of particles with energy equal to 1 up to random fluctuations that vanish with large $N$, and interactions given by a normal distribution with zero mean and variance 0.1. 
Then, we study the same problem but with $M_{kk}\sim\mathcal{N}(1,0.2)$. %, i.e., the variance of the diagonal elements is fixed at 0.2 independently of $N$. 
The results are depicted respectively in Fig.~\ref{fig:randomHam}(a) and~(b): only in the first case we obtain energy equipartition for $N\gg1$, as also captured by the behavior of $n_\text{eff}$ shown in Fig.~\ref{fig:randomHam}(c). Note that, while $M$ may have negative eigenvalues, we can set them positive by shifting $M$ through $\delta\mathbb{I}$ with $\delta=\mathcal{O}(\sqrt{N})$ \cite{Livan2018}. We have also verified that the equipartition emerges for the non-Wigner matrix $M M^\dagger$, which is positive-semidefinite by construction and displays eigenvector delocalization for large $N$. 

\bibliography{draftPRLbiblio}

\let\oldaddcontentsline\addcontentsline% Store \addcontentsline
\let\addcontentsline\oldaddcontentsline% Restore \addcontentsline
%\bibliographystyle{apsrev4-2}
%\bibliographystyle{vancouver}
%\bibliographystyle{iopart-num}
%\end{multicols}
\newpage
\pagebreak
\newpage
\widetext
\begin{center}
\textbf{\large Supplemental Material: Thermalization is typical in large classical and quantum harmonic systems}
\end{center}
%%%%%%%%%% Merge with supplemental materials %%%%%%%%%%
%%%%%%%%%% Prefix a "S" to all equations, figures, tables and reset the counter %%%%%%%%%%
\setcounter{equation}{0}
\setcounter{figure}{0}
\setcounter{table}{0}
\makeatletter
\renewcommand{\thefigure}{S\arabic{figure}}
\renewcommand{\theequation}{S\arabic{equation}}
\tableofcontents

%% ACHTUNG! References to the equations in the main text in the section titles need to be UPDATED MANUALLY

\section{Calculation of the dynamical averages in Eqs.~(5) and~(13)}
\subsection{Randomly rotated modes}
We consider the energy of a rotated mode as given by Eq.~(4):
\begin{equation}\label{eqn:rotatedModeSM}
    \tilde{E}^{(rot)}_k(t)=\mu_{kk}\tilde{o}_k^*(t)\tilde{o}_k(t) 
\end{equation}
We recall that $\mu_{kk'}=\sum_{j}C_{kj}C_{k'j}^*\omega_j$ and $o_k(t)=e^{-i\omega_k t}o_k(0)$, $o_k^*(t)=e^{i\omega_k t}o_k^*(0)$, were $\omega_k$ are the frequencies of the normal modes and the relation between $o_k$ and $\tilde{o}_k$ is expressed in Eq.~(3).  We want to compute its time average according to Eq.~\eqref{eqn:timeAv} and take the limit $t\rightarrow\infty$. The average is formally different in CM and QM, as in QM we need to introduce an additional average of the observable over the initial state. So, for the sake of clarity, let us perform the two calculations separately.

\subsubsection{Classical case}
In the classical case we use $o^*_k=z_k^*$, $o_k=z_k$ \cite{Strocchi1966}. Then,
\begin{equation}
\begin{split}
\label{eqn:derAvTimeRotClas}
    \overline{\tilde{E}^{(rot)}_k}_t &= \frac{1}{t}\int_0^t ds \, \mu_{kk} \sum_{j,j'} C_{kj}z_j(s) C_{kj'}^*z_{j'}^*(s)\\
    &=\frac{\mu_{kk}}{t}\sum_{j,j'} C_{kj}C_{kj'}^* z_j(0)z_{j'}^*(0) \int_0^t ds\, e^{-i(\omega_j-\omega_{j'})s}\\
    &=\mu_{kk}\sum_{j,j':j\neq j'} C_{kj}C_{kj'}^* z_j(0)z_{j'}^*(0) \frac{e^{-i(\omega_j-\omega_{j'})t}-1}{(\omega_j'-\omega_{j})t}+\mu_{kk}\sum_j \abs{C_{kj}}^2 \abs{z_j(0)}^2.\\
\end{split}
\end{equation}
Here, we have assumed that the frequencies of the normal modes are non-degenerate.
The first term clearly vanishes in the limit $t\rightarrow\infty$. On the contrary, the second term is independent of time. Then, inserting the expression for $\mu_{kk}$,
\begin{equation}
\label{eqn:infiniteEnergy}
    \overline{\tilde{E}^{(rot)}_k}_\infty = \sum_{j,j'} \omega_{j'} \abs{C_{kj'}}^2  \abs{C_{kj}}^2 \abs{z_j(0)}^2,
\end{equation}
which is Eq.~\eqref{eqn:Energy_Rot_modes} of the main text with $p_{0,j}=\abs{z_j(0)}^2$.

\subsubsection{Quantum case}
In the quantum case we replace $o_k^*=a_k^\dagger$, $o_k=a_k$, where $a_k$ are fermionic or bosonic operators. The time average is written as
\begin{equation}
\begin{split}
    \overline{\tilde{E}^{(rot)}_k}_t &= \frac{1}{t}\int_0^t ds \, \Tr[\mu_{kk} \sum_{j,j'} C_{kj'}^*a_{j'}^\dagger(s) C_{kj}a_j(s) \rho_0] \\
    &=\frac{\mu_{kk}}{t}\sum_{j,j'} C_{kj}C_{kj'}^* \Tr[a_{j'}^\dagger a_j \rho_0] \int_0^t ds\, e^{-i(\omega_j-\omega_{j'})s}\\
    &=\mu_{kk}\sum_{j,j':j\neq j'} C_{kj}C_{kj'}^* \Tr[a_{j'}^\dagger a_j \rho_0] \frac{e^{-i(\omega_j-\omega_{j'})t}-1}{(\omega_j'-\omega_{j})t}+\mu_{kk}\sum_j \abs{C_{kj}}^2 \Tr[a_{j}^\dagger a_j \rho_0],\\
\end{split}
\end{equation}
where $\rho_0$ is the initial state of the dynamics. By using the same argument as for the classical case, in the limit $t\rightarrow\infty$ we recover Eq.~\eqref{eqn:Energy_Rot_modes} with $p_{0,j}=\Tr[a_{j}^\dagger a_j \rho_0]$.

\subsection{Random Hamiltonian}
In this section we focus on Eq.~\eqref{eqn:randomHam}, which is the Hamiltonian of a $N$-particle system in which the number of excitations is conserved:
\begin{equation}
    H = \sum_{j,k=1} M_{jk} \tilde{o}_j^* \tilde{o}_k = \sum_k \tilde{E}_k^{(loc)} + \sum_{j\neq k} M_{jk} \tilde{o}_j^* \tilde{o}_k,
\end{equation}
where $M^\dagger=M$ and we have defined the energy of each local mode as
\begin{equation}
    \tilde{E}_k^{(loc)}= M_{kk} \tilde{o}_k^* \tilde{o}_k.
\end{equation}

The matrix of the quadratic form in the Hamiltonian can be diagonalized through a transformation $C$, which we can choose as unitary $C^\dagger = C^{-1}$, such that
\begin{equation}
    D = C^{-1} M C.
\end{equation}
The columns of $C$ are the eigenvectors of $M$. The diagonal matrix $D$ contains the eigenfrequencies of the normal modes: $D_{jk}=\omega_j \delta_{jk}$.
The Hamiltonian can then be written in the basis of the normal modes as
\begin{equation}
    H = \sum_{j,k} \sum_{m,n} C_{jm} D_{mn} C^\dagger_{nk} \tilde{o}_j^* \tilde{o}_k=\sum_m \omega_m o_m^* o_m,
\end{equation}
where we have introduced the operators for the normal modes
\begin{equation}
    o_m = \sum_j C_{jm}^* \tilde{o}_j,\quad o_m^*=\sum_j C_{jm} \tilde{o}_j^*.
\end{equation}
We can quickly verify that these operators satisfy the commutation relations (Poisson brackets). Equivalently, $\tilde{o}_j=\sum_k C_{jk}o_k$ and $\tilde{o}_j^* = \sum_k C_{jk}^* o_k^*$.

We can now write the observables we are interested in, namely the local energies $\tilde{E}_k^{(loc)}$, as a function of the normal modes:
\begin{equation}
    \tilde{E}_k^{(loc)} = \sum_{j,j'} M_{kk}  C_{kj'}^*C_{kj} o_{j'}^* o_{j}.
\end{equation}
Finally, we compute the time average of this quantity at infinite time, which can be written as
\begin{equation}
\label{eqn:derAvTimeRandomHam}
    \overline{\tilde{E}_k^{(loc)}}_\infty = \sum_{j,j'} M_{kk} C_{kj} C_{kj'}^* \lim_{t\rightarrow\infty} \frac{1}{t}\int_0 ^t ds  \, o_{j'}^*(s) o_{j}(s),
\end{equation}
where we remind that in the quantum case we have to insert an additional average over the initial state of the system $\rho_0$. Eq.~\eqref{eqn:derAvTimeRandomHam} is formally equivalent to Eq.~\eqref{eqn:derAvTimeRotClas}, so we immediately recover Eq.~\eqref{eqn:Energy_Loc_modes} under the assumption of non-degenerate $\omega_j$. Note that, in the quantum case for fermions, these calculations are similar for instance to \cite{Magan2016,Lydzba2023}, although our focus is different.

\section{Statistical properties of a random vector on the sphere}
\mar{The $k$th row of a random rotation $C$ distributed uniformly over the sphere can be described by \cite{GeorgeMarsaglia72,ORourke2016}
\begin{equation}
\label{eqn:randomVectorDistribution}
    C_{k\cdot} = \left(\frac{\xi_1^{(k)}}{\sqrt{\sum_j \left(\xi_j^{(k)}\right)^2}},\ldots,\frac{\xi^{(k)}_N}{\sqrt{\sum_j \left(\xi_j^{(k)}\right)^2}}\right),
\end{equation}
where $\xi_j^{(k)}$ are independently and identically distributed (iid) standard normal variables.
We are interested in the moments of the products of $\abs{C_{jk}}^2$. These can be computed through integrals in spherical coordinates on the $N$-dimensional sphere, and their expressions are well-known in the literature. For completeness, in this section we provide the derivation of all the relevant moments of $\abs{C_{jk}}^2$, as we will use them repetitively in the proofs of our results.}

We will use the following known integrals:
\begin{equation}
\label{eqn:listIntegrals}
    \begin{split}
        &\int_0^\infty dr\, \frac{r^{n-1}}{(2\pi)^{n/2}} e^{-r^2/2} = \frac{\pi^{-n/2}}{2}\Gamma\left(\frac{n}{2}\right),\\
        &\int_0^\pi d\varphi\, \cos^m(\varphi) \sin^n(\varphi)= \frac{\pi^{1/2}\,m!}{4^{m/2}(m/2)!}\frac{\Gamma\left(\frac{n+1}{2}\right)}{\Gamma\left(\frac{n+m}{2}+1\right)},\quad m\text{ even},\\
        &\int_0^\pi d\varphi\, \sin^n(\varphi)= \pi^{1/2}\frac{\Gamma\left(\frac{n+1}{2}\right)}{\Gamma\left(\frac{n}{2}+1\right)}.\\
    \end{split}
\end{equation}
We recall that the volume element on the $N$-sphere is expressed by
\begin{equation}
    d^NV = r^{N-1} \sin^{N-2}(\varphi_1) \sin^{N-3}(\varphi_2)\ldots\sin(\varphi_{N-2})\, dr d\varphi_1\ldots d\varphi_{N-1},
\end{equation}
and the spherical coordinates are given by
\begin{equation}
    \begin{split}
        &x_1= r \cos\varphi_1\\
        &x_2= r \cos\varphi_2 \sin\varphi_1\\
        &x_3= r \cos\varphi_3 \sin\varphi_2\sin\varphi_1\\
        &\ldots\\
        &x_{N-1}= r \cos\varphi_{N-1} \prod_{j=1}^{N-2}\sin\varphi_j\\
        &x_{N}= r \prod_{j=1}^{N-1}\sin\varphi_j,\\
    \end{split}
\end{equation}
where $r\in[0,\infty)$, $\varphi_j=[0,\pi]$ for $j=1,\ldots,N-2$ and $\varphi_{N-1}=[0,2\pi)$.

Then, we obtain
\begin{equation}
    \begin{split}
        \mathbb{E}[\abs{C_{jk}}^2]&=\int_0^\infty dr\, \frac{r^{N-1}}{(2\pi)^{N/2}} e^{-r^2/2}\int_0^\pi d\varphi_1\, \cos^2(\varphi_1) \sin^{N-2}(\varphi_1) \prod_{j'=1}^{N-3}\int_0^\pi d\varphi_{N-j'-1}\, \sin^{j'}(\varphi_{N-j'-1})\int_0^{2\pi}d\varphi_{N-1}\\
        &=\frac{1}{2}\frac{\Gamma\left(\frac{N}{2}\right)\Gamma\left(\frac{N-1}{2}\right)}{\Gamma\left(\frac{N}{2}+1\right)}\prod_{j'=1}^{N-3}\frac{\Gamma\left(\frac{j'+1}{2}\right)}{\Gamma\left(\frac{j'}{2}+1\right)}=\frac{1}{2}\frac{\Gamma\left(\frac{N}{2}\right)}{\Gamma\left(\frac{N}{2}+1\right)}= \frac{1}{N},
    \end{split}
\end{equation}
which is the result we expect for a random vector on the sphere. We have used the fact that
\begin{equation}
    \prod_{j=1}^{n}\frac{\Gamma\left(\frac{j+1}{2}\right)}{\Gamma\left(\frac{j}{2}+1\right)}=\frac{1}{\Gamma\left(\frac{n}{2}+1\right)},
\end{equation}
and the property of the Gamma function $\Gamma(z+1)=z\Gamma(z)$.
Analogously,
\begin{equation}
    \begin{split}
        \mathbb{E}[\abs{C_{jk}}^4]&=\int_0^\infty dr\, \frac{r^{N-1}}{(2\pi)^{N/2}} e^{-r^2/2}\int_0^\pi d\varphi_1\, \cos^4(\varphi_1) \sin^{N-2}(\varphi_1) \prod_{j'=1}^{N-3}\int_0^\pi d\varphi_{N-j'-1}\, \sin^{j'}(\varphi_{N-j'-1})\int_0^{2\pi}d\varphi_{N-1}\\
        &=\frac{3}{4}\frac{\Gamma\left(\frac{N}{2}\right)\Gamma\left(\frac{N-1}{2}\right)}{\Gamma\left(\frac{N}{2}+2\right)}\frac{1}{\Gamma\left(\frac{N-1}{2}\right)}=\frac{3}{4}\frac{\Gamma\left(\frac{N}{2}\right)}{\Gamma\left(\frac{N}{2}+2\right)}= \frac{3}{N(N+2)}.
    \end{split}
\end{equation}
\begin{equation}
\label{eqn:Proof8MomentRot}
    \begin{split}
        \mathbb{E}[\abs{C_{jk}}^8]&=\int_0^\infty dr\, \frac{r^{N-1}}{(2\pi)^{N/2}} e^{-r^2/2}\int_0^\pi d\varphi_1\, \cos^8(\varphi_1) \sin^{N-2}(\varphi_1) \prod_{j'=1}^{N-3}\int_0^\pi d\varphi_{N-j'-1}\, \sin^{j'}(\varphi_{N-j'-1})\int_0^{2\pi}d\varphi_{N-1}\\
        &=\frac{105}{16}\frac{\Gamma\left(\frac{N}{2}\right)\Gamma\left(\frac{N-1}{2}\right)}{\Gamma\left(\frac{N}{2}+2\right)}\frac{1}{\Gamma\left(\frac{N-1}{2}\right)}=\frac{105}{16}\frac{\Gamma\left(\frac{N}{2}\right)}{\Gamma\left(\frac{N}{2}+4\right)}= \frac{105}{N^4}+\mathcal{O}\left(\frac{1}{N^5}\right),
    \end{split}
\end{equation}
where we are considering $N\gg 1$.
Moreover, for $k\neq k'$ we have
\begin{equation}
    \begin{split}
        \mathbb{E}[\abs{C_{jk}}^2\abs{C_{jk'}}^2]=&\int_0^\infty dr\, \frac{r^{N-1}}{(2\pi)^{N/2}} e^{-r^2/2}\int_0^\pi d\varphi_1\, \cos^2(\varphi_1) \sin^{N}(\varphi_1)\int_0^\pi d\varphi_2\, \cos^2(\varphi_2) \sin^{N-3}(\varphi_2) \\
        &\prod_{j'=1}^{N-4}\int_0^\pi d\varphi_{N-j'-1}\, \sin^{j'}(\varphi_{N-j'-1})\int_0^{2\pi}d\varphi_{N-1}\\
        =&\frac{1}{4}\frac{\Gamma\left(\frac{N}{2}\right)\Gamma\left(\frac{N+1}{2}\right)\Gamma\left(\frac{N}{2}-1\right)}{\Gamma\left(\frac{N}{2}+2\right)\Gamma\left(\frac{N+1}{2}\right)}\frac{1}{\Gamma\left(\frac{N}{2}-1\right)}=\frac{1}{4}\frac{\Gamma\left(\frac{N}{2}\right)}{\Gamma\left(\frac{N}{2}+2\right)}= \frac{1}{N(N+2)}.
    \end{split}
\end{equation}
The key message to extract from these expectation values is that $\abs{C_{jk}}^2$ and $\abs{C_{jk'}}^2$ \textit{are independent} for $N\gg 1$ if $k\neq k'$. In contrast $E[\abs{C_{jk}}^4]\neq E[\abs{C_{jk}}^2]^2$.

Moreover, the four-point correlation function with $k\neq k'\neq m \neq m'$ reads
\begin{equation}
\label{eqn:ProofsFourPointCorr}
    \begin{split}
        \mathbb{E}[\abs{C_{jk}}^2\abs{C_{jk'}}^2\abs{C_{jm}}^2\abs{C_{jm'}}^2]=&\int_0^\infty dr\, \frac{r^{N-1}}{(2\pi)^{N/2}} e^{-r^2/2}\int_0^\pi d\varphi_1\, \cos^2(\varphi_1) \sin^{N+4}(\varphi_1)\int_0^\pi d\varphi_2\, \cos^2(\varphi_2) \sin^{N+1}(\varphi_2) \\
        \int_0^\pi d\varphi_3\, \cos^2(\varphi_3) \sin^{N-2}(\varphi_3)&\int_0^\pi d\varphi_4\, \cos^2(\varphi_4) \sin^{N-5}(\varphi_4) \prod_{j'=1}^{N-6}\int_0^\pi d\varphi_{N-j'-1}\, \sin^{j'}(\varphi_{N-j'-1})\int_0^{2\pi}d\varphi_{N-1}\\
        =&\frac{1}{16}\frac{\Gamma\left(\frac{N}{2}\right)}{\Gamma\left(\frac{N}{2}+4\right)}= \frac{1}{N^4}+\mathcal{O}\left(\frac{1}{N^5}\right),
    \end{split}
\end{equation}
while for $k\neq m\neq m'$ we have
\begin{equation}
\label{eqn:ProofsMixedThreePointCorr}
    \begin{split}
        \mathbb{E}[\abs{C_{jk}}^4\abs{C_{jm}}^2\abs{C_{jm'}}^2]=&\int_0^\infty dr\, \frac{r^{N-1}}{(2\pi)^{N/2}} e^{-r^2/2}\int_0^\pi d\varphi_1\, \cos^4(\varphi_1) \sin^{N+2}(\varphi_1)\int_0^\pi d\varphi_2\, \cos^2(\varphi_2) \sin^{N-1}(\varphi_2) \\
        &\int_0^\pi d\varphi_3\, \cos^2(\varphi_3) \sin^{N-4}(\varphi_3)\prod_{j'=1}^{N-5}\int_0^\pi d\varphi_{N-j'-1}\, \sin^{j'}(\varphi_{N-j'-1})\int_0^{2\pi}d\varphi_{N-1}\\
        =&\frac{3}{16}\frac{\Gamma\left(\frac{N}{2}\right)}{\Gamma\left(\frac{N}{2}+4\right)}= \frac{3}{N^4}+\mathcal{O}\left(\frac{1}{N^5}\right).
    \end{split}
\end{equation}

\mar{Finally, note that the expectation value of the product of two different elements of the matrix $C$ is always zero. For $k\neq k'$,
\begin{equation}
\begin{split}
    \mathbb{E}[C_{jk}C_{jk'}]=& \int_0^\infty dr\, \frac{r^{N-1}}{(2\pi)^{N/2}} e^{-r^2/2}\int_0^\pi d\varphi_1\, \cos(\varphi_1) \sin^{N-1}(\varphi_1)\int_0^\pi d\varphi_2\, \cos(\varphi_2) \sin^{N-3}(\varphi_2) \\
        &\prod_{j'=1}^{N-4}\int_0^\pi d\varphi_{N-j'-1}\, \sin^{j'}(\varphi_{N-j'-1})\int_0^{2\pi}d\varphi_{N-1}=0,
\end{split}
\end{equation}
because $\int_0^\pi d\varphi\,\cos(\varphi) \sin^{N-1}(\varphi)=\frac{\sin^N(\pi)}{N}-\frac{\sin^N(0)}{N}=0.$
}

\section{Derivation of Eqs.~(6),~(8),~(9), and~(10)}
\label{sec:equivalence}

\subsection{Classical case}
\mar{In this section, for simplicity we refer to $E$ as the total energy in the microcanonical ensemble.} Let $\mu_N(\mathbf{z},\mathbf{z}^*)$ be the classical microcanonical distribution
\begin{equation}
\mu_N(\mathbf{z},\mathbf{z}^*)=\frac{1}{Z(E)}\delta(E-H(\mathbf{z},\mathbf{z}^*)).
\end{equation}
The average values of the populations of the normal modes are
\begin{align}
\label{eqn:classicalPopMicro}
\langle |z_k|^2\rangle =
	\int \prod_{n=1}^{N}\mathrm{d}z_{n}\mathrm{d}z_{n}^{*}\, |z_{k}|^{2}\,\frac{\delta\left (E- H\right )}{Z(E)} =  \int_{\varepsilon-\imath\infty}^{\varepsilon+\imath\infty}\frac{\mathrm{d}y}{2\,\pi\imath}
 \,e^{y\,E}\int \prod_{n=1}^{N}\mathrm{d}z_{n}\mathrm{d}z_{n}^{*}\, |z_{k}|^{2}\,
 \,\frac{e^{-y\,\sum_{l}^{N} \omega_{l} z_{l} z_{l}^{*}}}{Z(E)}=\frac{E}{N}\frac{1}{\omega_k}\,.
\end{align}
\mar{Moreover,
\begin{equation}
    \langle \abs{z_k}^4\rangle=   \int_{\varepsilon-\imath\infty}^{\varepsilon+\imath\infty}\frac{\mathrm{d}y}{2\,\pi\imath}
 \,e^{y\,E}\int \prod_{n=1}^{N}\mathrm{d}z_{n}\mathrm{d}z_{n}^{*}\, |z_{k}|^{4}\,
 \,\frac{e^{-y\,\sum_{l}^{N} \omega_{l} z_{l} z_{l}^{*}}}{Z(E)} = 2\langle \abs{z_k}^2\rangle^2. 
\end{equation}
Therefore, the variance of the population of a single normal mode in the microcanonical ensemble is equal to the square of the mean value, and the fluctuations in the populations of the normal modes do not vanish in the thermodynamic limit.} 

Let us now consider the energy of the $k$th rotated mode $\tilde{E}^{(rot)}_k$ introduced in Eq.~\eqref{eqn:Energy_Rot_modes}. We write it as $\tilde{E}^{(rot)}_k=M_{kk} \abs{\tilde{z}_k}^2$, with $M_{kk}=\sum_j \abs{C_{kj}}^2\omega_j$. In this way, we are formally using the same expression for both $\tilde{E}^{(rot)}_k$ and the energy of a local particle in the random Hamiltonian $\tilde{E}^{(loc)}_k$ defined in Eq.~\eqref{eqn:randomHam}. Therefore, all the results in this section will be valid for both scenarios.  
Let us now consider the long-time average of this energy $O_k^{(rot)}$, defined in Eq.~\eqref{eqn:Energy_Rot_modes}:
\begin{equation}
    O_k^{(rot)} = M_{kk}\sum_j C_{kj} \abs{z_j(0)}^2.
\end{equation}
We use Eq.~\eqref{eqn:classicalPopMicro} to draw the populations $\abs{z_j(0)}^2$ in the above equation from the microcanonical ensemble, and we get
\begin{equation}
\langle O_k^{(rot)}\rangle=\frac{E}{N}M_{kk}\sum_{j}\frac{|C_{kj}|^2}{\omega_j}.
\label{eq:ensemble_average_micro}
\end{equation}

Next, we compute the microcanonical expectation value of $\tilde{E}_k^{(rot)}$:
\begin{align}
	\int \prod_{n=1}^{N}\mathrm{d}\tilde{z}_{n}\mathrm{d}\tilde{z}_{n}^{*}\, |\tilde{z}_{k}|^{2}\,\frac{\delta\left (E- H\right )}{Z(E)} = \int_{\varepsilon-\imath\infty}^{\varepsilon+\imath\infty}\frac{\mathrm{d}y}{2\,\pi\imath}
 \,e^{y\,E}\int \prod_{n=1}^{N}\mathrm{d}\tilde{z}_{n}\mathrm{d}\tilde{z}_{n}^{*}\, |\tilde{z}_{k}|^{2}\,
 \,\frac{e^{-y\,\sum_{j,j'} M_{jj'} \tilde{z}_{j} \tilde{z}_{j'}^{*}}}{Z(E)}\overset{N\gg1}{\to} \frac{E}{N}(M^{-1})_{kk},
\end{align}
with the matrix $M=C diag(\omega_j) C^{-1}$. 
We thus conclude: 
\begin{align}
\langle \tilde{E}_k^{(rot)} \rangle = \langle O_k^{(rot)} \rangle.
\end{align}
Moreover, as we will show shortly, for typical initial conditions $O_k^{(rot)}\approx \langle \tilde{E}_k^{(rot)} \rangle$ with high probability for large $N$. 
To this aim, we have to prove that the fluctuations are small compared to the average value, i.e. \begin{equation}
\frac{\langle (O_k^{(rot)})^2 \rangle-\langle \tilde{E}_k^{(rot)} \rangle^2}{\langle \tilde{E}_k^{(rot)} \rangle^2}\overset{N\gg1}{\to} 0
\label{eq:typicalitySM}
\end{equation}
The quantity $\langle (O_k^{(rot)})^2 \rangle$ is
\begin{align}
\langle (O_k^{(rot)})^2 \rangle 
 &= M_{kk}^2 \sum_{j,j'} |C_{kj}|^2|C_{kj'}|^2 \int_{\varepsilon-\imath\infty}^{\varepsilon+\imath\infty}\frac{\mathrm{d}y}{2\,\pi\imath}
 \,e^{y\,E}\int \prod_{n=1}^{N}\mathrm{d}z_{n}\mathrm{d}z_{n}^{*}\, |z_{j}|^{2}|z_{j^\prime}|^{2}\,
 \,\frac{e^{-y\,\sum_{l}^{N} \omega_{l} z_{l} z_{l}^{*}}}{Z(E)}\overset{N\gg1}{\to}\nonumber\\ 
 &\overset{N\gg1}{\to}\frac{E^2}{N^2}M_{kk}^2\left[2\sum_k \frac{|C_{kj}|^4}{\omega_j^2}+\sum_{j\neq j^\prime}\frac{|C_{kj}|^2|C_{kj'}|^2}{\omega_j \omega_{j^\prime}}\right]=\langle \tilde{E}_k^{(rot)} \rangle^2 +  \frac{E^2}{N^2}M_{kk}^2\sum_j \frac{|C_{kj}|^4}{\omega_j^2}
\end{align}
Thus, the variance of $O_k^{(rot)}$ is 
\begin{align}
\langle (O_k^{(rot)})^2 \rangle 
 -\langle \tilde{E}_k^{(rot)} \rangle^2 =  \frac{E^2}{N^2}M_{kk}^2\sum_j \frac{|C_{kj}|^4}{\omega_j^2}= \frac{E^2}{N^2}M_{kk}^2R_N
\end{align}
where we have defined $R_N=\sum_j \frac{|C_{kj}|^4}{\omega_j^2}$. Note that the constants $\frac{E}{N}$ and $M_{kk}$ drop out when divided by $\langle \tilde{E}_k^{(rot)} \rangle^2$. Thus, considering that $\langle \tilde{E}_k^{(rot)} \rangle=O(1)$, the problem reduces to show that $R_N\approx O\left(\frac{1}{N}\right)$ for $N\gg 1$. Assuming that $C_{kj}$ are vectors uniformly distributed on the unit sphere, one has:

%After some algebra, the quantity $\langle (O_k^{(rot)})^2 \rangle$ reads

%\begin{align}
%\langle (O_k^{(rot)})^2 \rangle 
% =\langle \tilde{E}_k^{(rot)} \rangle^2 +  \frac{E^2}{N^2}M_{kk}^2\sum_k \frac{|C_{kj}|^4}{\omega_k^2}=\langle \tilde{E}_k^{(rot)} \rangle^2 +  \frac{E^2}{N^2}M_{kk}^2R_N
%\nonumber
%\end{align}
%where we have defined $R_N=\sum_k \frac{|C_{kj}|^4}{\omega_k^2}$. 
\begin{align}
    &\mathbb{E}\left[R_N\right]=\frac{3}{N^2}\sum_j \frac{1}{\omega_j^2}+O\left(\frac{1}{N^3}\right)\\
    &\mathbb{E}\left[R_N^2\right]=\frac{96}{N^4}\sum_j\frac{1}{\omega_j^4}+\frac{9}{N^4}\sum_{j,j'} \frac{1}{\omega_j^2\omega_{j^\prime}^2}+O\left(\frac{1}{N^5}\right)\\
    &\frac{\text{Var}\left[R_N\right]}{\mathbb{E}\left[R_N\right]^2}=\frac{96\sum_j\frac{1}{\omega_j^4}}{\sum_{j,j'} \frac{1}{\omega_j^2\omega_{j^\prime}^2}}+O\left(\frac{1}{N}\right)
\end{align}
The behavior of $\frac{\text{Var}\left[R_N\right]}{\mathbb{E}\left[R_N\right]^2}$ depends on the assumptions on the distribution of eigenvalues. Focusing on systems with the property
\begin{equation}
    \lim_{N\to\infty}\frac{1}{N}\sum_k \frac{1}{\omega_k^n}<\infty \qquad \text{for } n=1,2,4%\in\mathbb{N}
    \label{eq:assumptions_omega}
\end{equation}
the ratio between variance and mean vanishes for large $N$, i.e.
\begin{align}
    &\frac{\text{Var}\left[R_N\right]}{\mathbb{E}\left[R_N\right]^2}=O\left(\frac{1}{N}\right)\\
    &R_N\approx\mathbb{E}\left[R_N\right]=O\left(\frac{1}{N}\right)\,.
\end{align}
%showing that Eq.\ref{eq:typicality} holds since Eq.~\ref{eq:assumptions_omega} implies $\langle O_i\rangle=O(1)$.
An alternative approach consists in assuming the existence of an infrared cutoff, i.e. $\omega_k\ge \omega_m\quad \forall k \in [1,N]$. This allow us to write $R_N \le \frac{1}{\omega_m^2}\sum_k|C_{kj}|^4\approx O\left(\frac{1}{N}\right)$ and adopting the same strategy as before we show the typicality of the result.

\subsubsection{Restricted ``microcanonical" distribution}\label{appendix:micro_restricted}

The ``restricted" microcanonical ensemble is almost equivalent to the microcanonical one except that only $N_S=\alpha N$ eigenmodes are excited. The distribution $\mu_N^S(\mathbf{z},\mathbf{z}^*)$ can be written as 
\begin{equation}
\mu_N^S(\mathbf{z},\mathbf{z}^*)=\frac{1}{Z_S(E)}\delta(E-H_S(\mathbf{z},\mathbf{z}^*))\prod_{k\in S^c}\delta(z_k)\delta(z_k^*)
\end{equation}
where $S^c$ is the complement of set $S$,
\begin{align}
H_S= \sum_{k\in S} \omega_k z_{k} z_{k}^{*}
%\nonumber
\end{align}
and $Z_S(E)$ is the microcanonical partition function for an harmonic system with $N_S$ degrees of freedom. 
Computations identical to those in the previous section lead to the following expressions for the amplitude of eigenmodes and the average of single-particle energy:
\begin{equation}
\langle |z_k|^2\rangle_S=\begin{cases}\frac{E}{N_S}\frac{1}{\omega_k}&\text{ if } k\in S\\
 0 &\text{ if } k\in S^c
\end{cases}
\end{equation}

\begin{equation}
\langle O_k^{(rot)}\rangle_S=\langle \tilde{E}_k^{(rot)}\rangle_S=\frac{E}{N_S}M_{kk}\sum_{j\in S}\frac{|C_{kj}|^2}{\omega_j}
\end{equation}
Moreover, by replacing the summation over $k\in[1,N]$ with a summation over $j\in S$ in the formula for $\langle O_k^{(rot)}\rangle$ one obtains
\begin{align}
\langle (O_k^{(rot)})^2 \rangle_S 
 =\langle \tilde{E}_k^{(rot)} \rangle^2_S +  \frac{E^2}{N_S^2}M_{kk}^2\sum_{j\in S} \frac{|C_{kj}|^4}{\omega_j^2}=\langle \tilde{E}_k^{(rot)} \rangle^2_S +  \frac{E^2}{N_S^2}M_{kk}^2R_{N_S}.
%\nonumber
\end{align}

\begin{comment}
\begin{align}
    &\mathbb{E}\left[R_{N_S}\right]\approx\frac{3}{N^2}\sum_{k\in S} \frac{1}{\omega_k^2}
    \nonumber\\
    &\mathbb{E}\left[R_N^2\right]\approx\frac{96}{N^4}\sum_{k\in S}\frac{1}{\omega_k^4}+\frac{9}{N^4}\sum_{k,k^\prime\in S} \frac{1}{\omega_k^2\omega_{k^\prime}^2}\nonumber\\
    &\frac{\text{Var}\left[R_N\right]}{\mathbb{E}\left[R_N\right]^2}\approx\frac{96\sum_{k\in S}\frac{1}{\omega_k^4}}{\sum_{k,k^\prime\in S} \frac{1}{\omega_k^2\omega_{k^\prime}^2}}
\end{align}
\end{comment}
Assuming again
\begin{equation}
    \lim_{N_S\to\infty}\frac{1}{N_S}\sum_{k\in S} \frac{1}{\omega_k^n}<\infty \qquad \text{for } n=1,2,4,%\in\mathbb{N}
    \label{eq:assumptions_omega_restricted}
\end{equation}
one recovers
\begin{align}
    &\frac{\text{Var}\left[R_{N_S}\right]}{\mathbb{E}\left[R_{N_S}\right]^2}=O\left(\frac{1}{N_S}\right)\nonumber\\
    &R_{N_S}\approx\mathbb{E}\left[R_{N_S}\right]=O\left(\frac{1}{N}\right)
\end{align}
%showing that Eq.\ref{eq:typicality} holds since Eq.~\ref{eq:assumptions_omega} implies $\langle O_i\rangle=O(1)$. Finally, from $$\lim_{N\to\infty}\mu_N=\lim_{N\to\infty}\mu_N^S$$
%follows the typicality also with respect to the microcanonical ensemble. 
proving that $O_k^{(loc)}\approx \langle\tilde{E}_k^{(rot)}\rangle_S$. Finally, we want to show that typicality with respect $\mu_N^S$ implies typicality with respect to $\mu_N$. Regarding the average value, it is possible to prove that 
\begin{equation}
\langle \tilde{E}_k^{(rot)}\rangle_S=\frac{E}{N_S}M_{kk}\sum_{j\in S}\frac{|C_{kj}|^2}{\omega_j}\approx \frac{E}{N}\frac{M_{kk}}{N_S}\sum_{j\in S}{\frac{1}{\omega_j}}\overset{N\gg1}{\to}\langle \tilde{E}_k^{(rot)}\rangle,
\end{equation}
provided that 
\begin{equation}
    \frac{1}{N_S}\sum_{j\in S}{\frac{1}{\omega_j}}=\frac{1}{N}\sum_j{\frac{1}{\omega_j}} \text{ 
for $N\gg1$.}
\end{equation}

\subsection{Quantum case}% (DA SISTEMARE SULLA FALSARIGA DEL CASO PRECEDENTE)}

In the Segal-Bargman (or holomorphic) representation, see e.g. chapters~6 for Bosons and 7 for Fermions of \cite{ZiJu2010}, the Hamilton operator takes the form
\begin{align}
H=\sum_{i,j=1}^{N}M_{ij}\,z_{i} \partial_{z_{j}}.
\end{align}
Let us first focus on the Boson case. The change of variables 
\begin{align}
w_{i}=\sum_{j=1}^{N} z_{j}\,C_{ji}
\end{align}
specified by the unitary $C$ that diagonalizes $M$ also couches the Hamiltonian into diagonal form:
\begin{align}
H=\sum_{i=1}^{N}\omega_{i}\,w_{i} \partial_{w_{i}}.
\end{align}
We thus verify that the eigenvalues are 
\begin{align}
	h_{n_{1},\dots,n_{N}}=\sum_{i=1}^{N}n_{i}\omega_{i}.
\end{align} 
In order to compute the microcanonical partition function we need to compute the degeneracy in $[n_{1},\dots,n_{N}]$ of an eigenstate of fixed energy $E_\text{tot}$ see e.g. chapter~8 of \cite{GrNeSt1995}:
\begin{align}
	Z(E_\text{tot})=|\left\{ [n_{1},\dots,n_{N}]\,\big{|} E_\text{tot}= h_{n_{1},\dots,n_{N}}\right\}|.
\end{align} 
In the simplest case when all the eigenfrequencies are equal, 
\begin{align}
E_\text{tot}=\omega \sum_{i=1}^{N}n_{i}=\omega M,
\end{align}
the computation reduces to counting the number of ways to allocate $M$ indistingushable balls into $N$ urns:
\begin{align}
	Z(\omega M )=\frac{(N+M-1)!}{M!(N-1)!}.
\end{align}
Using Boltzmann's definition of the temperature we straighforwardly arrive at
\begin{align}
	E_\text{tot}=\frac{N\omega}{e^{\beta  \omega}-1}.
\end{align}
Further calculations become somewhat awkward in the microcanonical ensemble. We can, however, rely on the equivalence of statistical ensembles for $N\gg 1$. We thus resort to the canonical ensemble. It is straightforward to verify that for bosons
\begin{align}
	Z_{C}(\beta)=\operatorname{Tr}e^{-\beta\,\operatorname{H}}= \prod_{i=1}^{N}\sum_{n_{i}=1}^{\infty}e^{-n_{i}\beta\,\omega_{i}}
	=\prod_{i=1}^{N}\frac{1}{1-e^{-\beta\omega_{i}}}=\frac{1}{\det(\operatorname{1}-e^{-\beta\operatorname{M}})}.
\end{align}
In the holomorphic representation, this latter result corresponds to the Gaussian integral
\begin{align}
	\operatorname{Tr}e^{-\beta\operatorname{H}}=\int_{\mathbb{C}^{N}}\prod_{i=1}^{N}\frac{\mathrm{d}z\mathrm{d}\bar{z}}{2\,\pi} e^{-\sum_{i=1}^{N}z_{i}\bar{z}_{i} +\sum_{i,j=1}^{N}(e^{-\beta\operatorname{M}})_{ij} z_{i} \bar{z}_{j}} =\frac{1}{ \det\left(1-e^{-\beta\,\operatorname{M}}\right)}.
\end{align}
We avail us of this observation to evaluate
\begin{align}
	\operatorname{Tr}\left(e^{-\beta\operatorname{H}} \tilde{a}_{k}^{\dagger}\tilde{a}_{k}\right)
	=
	\int_{\mathbb{C}^{N}}\prod_{i=1}^{N}\frac{\mathrm{d}z\mathrm{d}\bar{z}}{2\,\pi} e^{-\sum_{i=1}^{N}z_{i}\bar{z}_{i} +\sum_{i,j=1}^{N}(e^{-\beta\operatorname{M}})_{ij} z_{i} \bar{z}_{j}}|z_{k}|^{2}
	=\frac{\left(\big{(}\operatorname{1}-e^{-\beta\operatorname{M}}\big{)}^{-1}\right)_{kk}}{
		\det\left(\operatorname{1}-e^{-\beta\,\operatorname{M}}\right)},
\end{align}
whence finally
\begin{align}
		\frac{1}{Z_{C}(\beta)}\operatorname{Tr}\left(e^{-\beta\,\operatorname{H}} \tilde{a}_{k}^{\dagger}\tilde{a}_{k}\right)=\sum_{i=1}^{N} \frac{|\operatorname{C}_{ki}|^{2}}{1-e^{-\beta\omega_{i}}}.
\end{align}
From this result it is clear that the calculation of the expectation value and the variance of $O_k^{(rot)}$ in the quantum case proceeds as in the classical case provided we perform the replacement
\begin{align}
    \frac{(M^{-1})_{kk}}{\beta} \mapsto ((1-e^{-\beta M})^{-1})_{kk}
\end{align}
For fermions, the holomorphic formalism uses Grassman variables and the trace operation must be replaced by the ``antitrace'' (i.e.) antiperiodic boundary conditions. The upshot is that in calculations
\begin{align}
    \frac{(M^{-1})_{kk}}{\beta} \mapsto ((1+e^{-\beta M})^{-1})_{kk}.
\end{align}
With these provisos it is straightforward both in the Boson and Fermion cases to recover the same conclusions regarding Khinchin typicality as in classical statistical mechanics.

\mar{Using the canonical ensemble and the equivalence of the ensembles, we can also immediately write the microcanonical temperature $\beta$ as a function of the total energy and the mode frequencies. In particular, $\beta$ is implicitly defined by
\begin{equation}
    E_\text{tot} = \sum_j \frac{\omega_j}{e^{\beta\omega_j}\pm 1},
\end{equation}
respectively for fermions and bosons. Moreover, we can also immediately calculate the variance of the populations of the normal modes $p_{0,j}$ in the microcanonical ensemble:
\begin{equation}
    \langle p_{0,j}^2\rangle -\langle p_{0,j} \rangle^2 = \bar{n}(1\mp \bar{n}),
\end{equation}
with $\bar{n}=1/(e^{\beta\omega_j}\pm 1)$, respectively for fermions and bosons.
}

\section{Statistical properties of the observables of interest in the case of roughly equal initial populations (Eqs.~(14),~(15),~(22),~(23),~(24))}

In what follows, we will frequently use the moments of the components of a random vector uniformly distributed on the sphere, as computed in the penultimate section, without explicitly referencing their formulas.

\subsubsection{Randomly rotated modes}
Let us consider the mean value of $O_k^{(rot)}$ in Eq.~\eqref{eqn:Energy_Rot_modes}, i.e., the energy of a randomly rotated mode, when the rotation $C$ is drawn randomly and uniformly over the sphere:
\begin{equation}
\label{eqn:ProofsMeanValRot}
    \begin{split}
        \mathbb{E}[O_k^{(rot)}] =& \sum_j \omega_j p_{0,j} \mathbb{E}[\abs{C_{kj}}^4]+ \sum_{\substack{j,j'\\j'\neq j}} \omega_jp_{0,j'} \mathbb{E}[\abs{C_{kj}}^2\abs{C_{kj'}}^2]= \frac{1}{N(N+2)}\left( \sum_j 3 \omega_j p_{0,j} + \sum_{\substack{j,j'\\j'\neq j}} \omega_jp_{0,j'}\right).
    \end{split}
\end{equation}
We define the average quantities $\omega_\text{av}=\sum_j\omega_j/N$ and $p_\text{av}=\sum_j p_{0,j}/N$. If we assume $\omega_j/\omega_\text{av}=\mathcal{O}(1)$ and  $p_{0,j}/p_\text{av}=\mathcal{O}(1)$ for all $j$, then $\sum_{\substack{j,j'\\j'\neq j}} \omega_jp_{0,j'}=\omega_\text{av}p_\text{av} \left(N^2+\mathcal{O}(1)\right)$. Moreover, $\sum_j\omega_jp_{0,j}=\mathcal{O}(N \omega_\text{av}p_\text{av})$, and in Eq.~\eqref{eqn:ProofsMeanValRot} it gets killed by the term going as $1/N^2$. Therefore, only the cross terms  over $j$ and $j'$ matter, and we obtain 
\begin{equation}
    \mathbb{E}[O_k^{(rot)}]= \omega_\text{av}p_\text{av}\left(1+\mathcal{O}\left(\frac{1}{N}\right)\right),
\end{equation}
which is the result in Eq.~\eqref{eqn:meanVrandomRotNonMicro}.
Next, we compute the variance:
\begin{equation}
\label{eqn:ProofsVarRot_crossterms}
    \begin{split}
        \Var[O_k^{(rot)}] =& \sum_{\substack{j,j',m,m'\\j'\neq j\neq m\neq m'}}  \omega_j \omega_m p_{0,j'} p_{0,m'} \mathbb{E}[\abs{C_{kj}}^2\abs{C_{kj'}}^2\abs{C_{km}}^2\abs{C_{km'}}^2]+ \mathcal{R} -\omega_\text{av}^2p_\text{av}^2+\mathcal{O}\left(\frac{\omega_\text{av}^2p_\text{av}^2}{N}\right).
    \end{split}
\end{equation}
The remainder $\mathcal{R}$ contains all the terms in which at least two of the indexes $j,j',m,m'$ are equal. Following a similar reasoning as for the mean value, all these terms are dominated by the most general cross-term in Eq.~\eqref{eqn:ProofsVarRot_crossterms} which is proportional to the four-point correlation function. Therefore, they get killed by the scaling $1/N^4$ and we can neglect all of them in the computation of the variance, i.e., $\mathcal{R}=\mathcal{O}\left(\omega_\text{av}^2p_\text{av}^2/N\right)$. Using Eq.~\eqref{eqn:ProofsFourPointCorr} and the standard assumption for the distribution of $\omega_j$ and $p_{0,j}$, the final result is then
\begin{equation}
\label{eqn:ProofsVarRot_finalresults}
    \begin{split}
        \Var[O_\infty^{(j)}] =\mathcal{O}\left(\frac{\omega_\text{av}^2p_\text{av}^2}{N}\right),
    \end{split}
\end{equation}
which proves Eq.~\eqref{eqn:finalScaling_rot}.

We are now able to understand why we need to require a uniform distribution (up to fluctuations that scale as $\mathcal{O}(1/N)$) of the eigenfrequencies and of the initial conditions from a mathematical perspective. Suppose, for instance, that $p_{0,j}=\delta_{jk^*}$ for a unique $k^*$, i.e., the initial state is localized on a single normal mode. Then, we have to insert a Dirac delta peaked on $k^*$ in each summation over the different initial conditions and, in particular, we cannot neglect the summations over $j'\neq j\neq m$ with $j'=m'$ in the remainder of Eq.~\eqref{eqn:ProofsVarRot_crossterms}, as they bring a contribution that is of the same order in $N$ as the one brought by the four-point cross term. These terms are proportional to the moment $\mathbb{E}[\abs{C_{kj'}}^4\abs{C_{km}}^2\abs{C_{kj}}^2]$, which is computed in Eq.~\eqref{eqn:ProofsMixedThreePointCorr}. Since the latter is proportional to $3/N^4$ instead of $1/N^4$, the terms of the order of $\mathcal{O}\left(\omega_\text{av}^2p_\text{av}^2/N^4\right)$ do not get cancelled out by $-\omega_\text{av}^2p_\text{av}^2$, and the final variance is of the same order as the square of the mean value, i.e., fluctuations around the mean value do not vanish for large $N$. The same argument would be valid if one frequency is much larger than all the rest, or if both the orders of magnitude of the frequencies and of the initial conditions would vary as a function of $j$ (in the latter case we would need to use the integral in Eq.~\eqref{eqn:Proof8MomentRot}, still getting to the same conclusions).% It is easy to understand that a similar behavior would be observed if the distribution of the frequencies and/or initial conditions over the normal modes would be restricted to a subset whose size divided by $N$ would become increasingly small for $N\rightarrow\infty$. 

\subsubsection{Random Hamiltonian}
We first consider the mean value of $O_k^{(loc)}$ in Eq.~\eqref{eqn:Energy_Loc_modes}, i.e., the energy of a local mode of the random Hamiltonian:
\begin{equation}
    \mathbb{E}[O_k^{(loc)}] = \mathbb{E}[M_{kk}] \sum_j \mathbb{E}[\abs{C_{kj}}^2] p_{0,j} = \mathbb{E}[M_{kk}] p_\text{av},
\end{equation}
where $p_\text{av}=\sum_j p_{0,j}/N$, and
we have assumed that the distribution of the diagonal elements of $M$ is independent of the eigenvectors. This is justified in the limit $N\gg 1$ if the eigenvectors get delocalized.

Next, let us calculate the variance of this quantity. We recall that, for two independent random variable $A$ and $B$, $\Var[AB]=\Var[A]\Var[B]+\Var[A]\mathbb{E}[B]^2+\Var[B]\mathbb{E}[A]^2$, therefore we can focus on the variance of the elements in the summation over $j$ in Eq.~\eqref{eqn:Energy_Loc_modes}. Specifically, we obtain
\begin{equation}
    \begin{split}
        \Var\left[\sum_j \abs{C_{kj}}^2 p_{0,j}\right] =& \sum_j \mathbb{E}[\abs{C_{kj}}^4] \left( p_{0,j}\right)^2 + \sum_{\substack{j,j'\\j\neq j'}}\mathbb{E}[\abs{C_{kj}}^2\abs{C_{kj'}}^2] p_{0,j}p_{0,j'} -p_\text{av}^2\\
        =& \frac{1}{N^2+2N}\left(3\sum_j \left( p_{0,j}\right)^2+\sum_{\substack{j,j'\\j\neq j'}} p_{0,j}p_{0,j'}\right)-p_\text{av}^2.
    \end{split}
\end{equation}
Once again, we observe that, assuming  $p_{0,j}/p_\text{av}=\mathcal{O}(1)$ for all $j$, $\sum_j \left( p_{0,j}\right)^2=\mathcal{O}(Np_\text{av}^2)$, while $\sum_{\substack{j,j'\\j\neq j'}} p_{0,j}p_{0,j'}=N^2p_\text{av}^2\left(1+\mathcal{O}(1/N^2)\right)$. Therefore,
\begin{equation}
        \Var\left[\sum_j \abs{C_{kj}}^2 p_{0,j}\right] = \mathcal{O}\left(\frac{p_\text{av}^2}{N}\right).
\end{equation}
Finally,
\begin{equation}
    \Var[O_k^{(loc)}]= \Var[M_{kk}]p_\text{av}^2+\mathcal{O}\left(\frac{\Var[M_{kk}]p_\text{av}^2}{N}\right)+\mathcal{O}\left(\frac{(\mathbb{E}[M_{kk}])^2p_\text{av}^2}{N}\right),
\end{equation}
which is the result in Eq.~\eqref{eqn:varLocMod}.

\section{Derivation of Eq.~(16)}
\label{sec:eqValue}
The total energy in the system is written as
\begin{equation}
    \label{eqn:totEnergy}
    E_\text{tot}=\sum_j \omega_j p_{0,j}.
\end{equation}
The difference between $E_\text{tot}/N$ and the equipartition value obtained in Eq.~\eqref{eqn:meanVrandomRotNonMicro}  is
\begin{equation}
    \Delta = \frac{1}{N}\sum_j \omega_j (p_{0,j} - p_\text{av}).
\end{equation}
Now let us assume that the initial populations are typical in the sense that they are (roughly) uniformly distributed among the normal modes. For instance, we can think of them as chosen randomly from the sphere of radius $R$. Equivalently, we can generate them as $p_{0,j}=c_r R/N$, where $c_r$ is a random number chosen uniformly from $[0,2)$. For all practical purposes, we can use the distribution in Eq.~\eqref{eqn:randomVectorDistribution} (eventually multiplied by the radius $R$) to describe a possible typical distribution of the population. The average of $p_{0,j}$ over the typical initial conditions is $\E_\text{ic}[p_{0,j}]=p_\text{av}$. If we now average $\Delta$ over the possible typical initial conditions, we obtain
\begin{equation}
    \E_\text{ic}[\Delta] = 0,\quad
    \Var[\Delta]_\text{ic} = \mathbb{E}_\text{ic}[\Delta ^2]= \frac{1}{N^2}\sum_{jk} \omega_j \omega_k (\mathbb{E}_\text{ic}[p_{0,j}p_{0,k}] - p_\text{av}^2) = \frac{\omega_\text{av}^2p_\text{av}^2}{N}(1+\mathcal{O}(1/N)).
\end{equation}
We have used the standard assumption $\omega_j/\omega_\text{av}=\mathcal{O}(1)$.
Then,
\begin{equation}
    \frac{\Var_\text{ic}[\Delta]}{\omega_\text{av}^2p_\text{av}^2} \rightarrow 0 \text{ for } N\gg1,
\end{equation}
and the equipartition value is always close to $E_\text{tot}/N$ for typical initial conditions and large $N$.

Note that the same result holds true also for the equipartition value obtained with a random Hamiltonian $M$, which is described by Eq.~\eqref{eqn:meanVlocal}, using $M_{kk}=\sum_j \abs{C_{kj}}^2 \omega_j$, if again we assume $\omega_j/\omega_\text{av}=\mathcal{O}(1)$ (note that this assumption is however not required to prove Eq.~\eqref{eqn:fluctuationsLocalMod}). 

\section{Derivation of the time variance (Eqs.~(18),~(19),~(20),~(21))}
\label{sec:eqValue}
\subsection{Microscopic observables}

\subsubsection{Classical case}
\mar{For CM the time average of the energies of the rotated mode  (see Eq.~\eqref{eqn:Energy_Rot_modes}) is
\begin{equation}
    \overline{\tilde{E}^{(rot)}_k}_\infty= \mu_{kk}\sum_j\abs{C_{kj}}^2\abs{z_j(0)}^2.
\end{equation}
We now compute
\begin{equation}
    \overline{(\tilde{E}^{(rot)}_k)^2}_\infty=\mu_{kk}^2\sum_{\alpha,\beta,\gamma,\delta}  C_{k\alpha}^* z_{\alpha}^*(0)C_{k\beta}z_\beta(0) C_{k\gamma}^*z_{\gamma}^*(0)C_{k\delta}z_\delta(0)\lim_{t\rightarrow\infty}\frac{1}{t}\int_0^t ds e^{-i(\omega_\beta-\omega_\alpha+\omega_\delta-\omega_\gamma)s}.
\end{equation}
From now on, including in the quantum case, we follow the same lines as in \cite{Lydzba2023}, which deals only with QM.}

\mar{In the derivation of Eq.~\eqref{eqn:Energy_Rot_modes} we made the assumption that there are no degenerate eigenfrequencies. Here, following the literature on ETH \cite{dalessio2015,Lydzba2023}, we make an even stronger assumption: there are no degeneracies in the \textit{gaps} between the eigenfrequencies. This is typically satisfied in complex systems. Under this assumption, the complex exponential in the above equation is different from zero only if: i) $\omega_\beta=\omega_\alpha$, $\omega_\gamma=\omega_\delta$; ii) $\omega_\beta=\omega_\gamma$, $\omega_\delta=\omega_\alpha$, and $\omega_\alpha\neq \omega_\beta$. Then,
\begin{equation}
\begin{split}
     \overline{(\tilde{E}^{(rot)}_k)^2}_\infty&= \mu_{kk}^2 \left(\sum_{\alpha,\beta}  \abs{C_{k\alpha}}^2 \abs{z_{\alpha}(0)}^2\abs{C_{k\beta}}^2 \abs{z_{\beta}(0)}^2 + \sum_{\alpha,\beta:\alpha\neq\beta} \abs{C_{k\alpha}}^2 \abs{z_{\alpha}(0)}^2\abs{C_{k\beta}}^2 \abs{z_{\beta}(0)}^2 \right)\\
     &\approx 2\mu_{kk}^2\sum_{\alpha,\beta}  \abs{C_{k\alpha}}^2 \abs{z_{\beta}(0)}^2\abs{C_{k\beta}}^2 \abs{z_{\alpha}(0)}^2 = 2 \left(\overline{\tilde{E}^{(rot)}_k}_\infty\right)^2. 
\end{split}
\end{equation}
In the above equation, we have used the fact that, for $N\gg 1$, adding a single element to the summation with $\alpha\neq\beta$ is negligible. 
Therefore, the variance of the time fluctuations is equal to the mean value, and the fluctuations over time do not vanish in the thermodynamic limit.}

\subsubsection{Quantum case}
\mar{
In the quantum case we perform an additional average of the observables over the initial state $\rho_0$, expressed as the trace $\Tr[O \rho_0]$. Then, the mean value is
\begin{equation}
    \overline{\tilde{E}^{(rot)}_k}_\infty= \mu_{kk}\sum_j\abs{C_{kj}}^2\Tr[a_j^\dagger a_j \rho_0].
\end{equation}
With the same assumptions as for the classical case, the time variance can be expressed as
\begin{equation}
\begin{split}
    \overline{(\tilde{E}^{(rot)}_k)^2}_\infty&= \mu_{kk}^2\sum_{\alpha,\beta,\gamma,\delta}  C_{k\alpha}^* C_{k\beta}\Tr[a_\alpha^\dagger a_\beta \rho_0] C_{k\gamma}^*C_{k\delta}\Tr[a_\gamma^\dagger a_\delta \rho_0]\lim_{t\rightarrow\infty}\frac{1}{t}\int_0^t ds e^{-i(\omega_\beta-\omega_\alpha+\omega_\gamma-\omega_\delta)s}\\
    &= \mu_{kk}^2\left( \sum_{\alpha,\beta} \abs{C_{k\alpha}}^2 \Tr[a^\dagger_\alpha a_\alpha\rho_0]\abs{C_{k\beta}}^2 \Tr[a^\dagger_\beta a_\beta\rho_0] + \sum_{\alpha\neq\beta}\abs{C_{k\alpha}}^2 \abs{C_{k\beta}}^2 \Tr[a^\dagger_\alpha a_\beta\rho_0] \Tr[a^\dagger_\beta a_\alpha\rho_0]\right).
\end{split}
\end{equation}
Note that, unlike for CM, the variance cannot be easily written as the square of the mean value anymore, as it depends on the cross-terms $\Tr[a^\dagger_\alpha a_\beta\rho_0]$. For instance, if the initial state is diagonal (no coherences in the Fock basis), then the variance is exactly zero. Moreover, following \cite{Lydzba2023} we can prove that, for fermions and hard-core bosons, the variance over the square of the mean value goes to zero for large $N$, independently of the initial conditions. Indeed, let us introduce the matrix defined by
\begin{equation}
    R_{\alpha\beta}= \Tr[a^\dagger_\beta a_\alpha\rho_0].
\end{equation}
We notice $R_{\alpha\beta}^*=R_{\beta\alpha}$. Then, 
\begin{equation}
    \sum_{\alpha\neq\beta}\abs{C_{k\alpha}}^2 \abs{C_{k\beta}}^2 \Tr[a^\dagger_\alpha a_\beta\rho_0] \Tr[a^\dagger_\beta a_\alpha\rho_0] = \sum_{\alpha\neq\beta}\abs{C_{k\alpha}}^2 \abs{C_{k\beta}}^2 \abs{R_{\alpha\beta}}^2\leq \max_\alpha{\abs{C_{k\alpha}}^4} \sum_\beta (R^2)_{\beta\beta}.
\end{equation}}

\mar{Next, it can be shown that the eigenvalues of $R$ are less or equal than 1, assuming we are dealing with fermions or hard-core bosons. Indeed, $R^\dagger=R$, so it can be diagonalized through a unitary transformation $M$, such that $M^\dagger RM =diag(\lambda_1,\ldots,\lambda_N)$. Then, 
\begin{equation}
    \lambda_\alpha = \sum_{\beta,\gamma} M_{\beta\alpha}^* R_{\beta\gamma} M_{\gamma\alpha} = \Tr\left[\sum_{\beta,\gamma}  M_{\gamma\alpha}a_\gamma^\dagger M_{\beta\alpha}^* a_\beta \rho_0\right]=\Tr[f_\alpha^\dagger f_\alpha\rho_0]\leq 1,
\end{equation}
where $f_\alpha= \sum_\beta M_{\beta\alpha}^* a_\beta$, and equivalently $f_\alpha^\dagger$, are some new well-defined fermionic (or hard-core bosonic) operators.
If this is the case, then
\begin{equation}
    \overline{(\tilde{E}^{(rot)}_k)^2}_\infty \leq \left(\overline{\tilde{E}^{(rot)}_k}_\infty\right)^2 + \max_\alpha{\abs{C_{k\alpha}}^4} \sum_\beta R_{\beta\beta} \leq \left(\overline{\tilde{E}^{(rot)}_k}_\infty\right)^2 + N \max_\alpha{\abs{C_{k\alpha}}^4}. 
\end{equation}
Following the same lines as for the derivations of the moments of the averages over the disorders in the previous sections, we can now replace $\abs{C_{k\alpha}}^4$ with $1/N^2$ for $N\gg1$. Therefore, the long-time variance of the microscopic observables in the quantum case vanishes for $N\gg 1$.}
\subsection{Macroscopic observables}
\mar{As we already know that in QM the time variance of even the microscopic observables vanishes, here for simplicity we consider only the classical case. We consider a macroscopic observable written as in Eq.~\eqref{eqn:macroObs}:
\begin{equation}
    A_M(t) = \sum_{k\in\mathsf{K}} \mu_{kk} \tilde{n}_k(t), \quad \tilde{n}_k(t)=\tilde{z}^*_k(t)\tilde{z}_k(t).
\end{equation}
$k$ runs over $N^*\gg 1$ rotated modes. Trivially,
\begin{equation}
    \overline{A_M}_\infty = \sum_{k\in\mathsf{K}} \overline{\tilde{E}^{(rot)}_k}_\infty = \sum_{k\in\mathsf{K}} \mu_{kk} \sum_{j=1}^N \abs{C_{kj}}^2 \abs{z_j(0)}^2.
\end{equation}
Next, we calculate the time variance (the indexes $k,k'$ run over $\mathsf{K}$, while the other indexes run from $1$ to $N$): 
\begin{equation}\label{eqn:varAM}
\begin{split}
    \overline{A_M^2}_\infty = &\sum_{k,k'} \mu_{kk} \mu_{k'k'} \lim_{t\rightarrow\infty}\frac{1}{t}\int_0^t ds \, \tilde{n}_k(s) \tilde{n}_{k'}(s)\\=& \sum_{k,k'} \mu_{kk} \mu_{k'k'} \sum_{j,j'}\sum_{r,r'} C_{kj}^* z_j^*(0) C_{kj'} z_{j'}(0) C_{k'r}^* z_r^*(0) C_{k'r'}  z_{r'}(0) \lim_{t\rightarrow\infty}\frac{1}{t}\int_0^t ds \, e^{-i(\omega_{j'}-\omega_j+\omega_{r'}-\omega_r)s}\\
     =& \sum_{k,k'} \mu_{kk} \mu_{k'k'} \left(\sum_j \abs{C_{kj}}^2 \abs{z_j(0)}^2 \sum_r \abs{C_{k'r}}^2\abs{z_r(0)}^2+ \sum_j C_{kj}^* C_{k'j} \abs{z_j(0)}^2\sum_{r:j\neq r} C_{k'r}^* C_{kr}\abs{z_r(0)}^2\right)\\
     = &\sum_{k,k'} \overline{\tilde{E}^{(rot)}_k}_\infty\; \overline{\tilde{E}^{(rot)}_{k'}}_\infty+ \sum_k  \overline{\tilde{E}^{(rot)}_k}_\infty^2 + \sum_{k,k':k\neq k'}\mu_{kk} \mu_{k'k'}   \sum_{j,r:j\neq r}C_{kj}^* C_{k'j} \abs{z_j(0)}^2 C_{k'r}^* C_{kr}\abs{z_r(0)}^2.
\end{split}
\end{equation}
The first term on the RHS of the above expression is equal to $\overline{A_M}_\infty^2$. }

\mar{
For simplicity, we now proceed in a non-rigorous way. We know that for $N\gg 1$ the value of $\sum_j\abs{C_{kj}}^2$ for a random $C$ gets peaked around $1$, and $\mu_{kk}$ gets peaked around $\omega_\text{av}$ (now to be computed with respect to the modes in $\mathsf{K}$). Then, for $N\gg 1$, the first term in the r.h.s. of the last equation reads
\begin{equation}\label{eqn:amdouble}
    \overline{A_M}_\infty^2 \approx  \left(\frac{N^*}{N}\right)^2 \omega_\text{av}^2 \sum_{j,j'} \abs{z_j(0)}^2\abs{z_{j'}(0)}^2.
\end{equation}
The error in the above equation scales as $\mathcal{O}(1/N^*)$, which absorbs also the contributions going as $\mathcal{O}(1/N)$. Next,
\begin{equation}
    \sum_k \overline{\tilde{E}^{(rot)}_k}_\infty^2 \approx \frac{N^*}{N^2}\omega_\text{av}^2  \sum_{j,j'} \abs{z_j(0)}^2\abs{z_{j'}(0)}^2 \ll \overline{A_M}_\infty^2.
\end{equation}
Finally, we obviously obtain $\E[ C_{kj}^* C_{k'j}]=0$ for $k\neq k'$, where $\E$ is the average over the disorder. Moreover, 
\begin{equation}
   \sum_{j,j'} \E[C_{kj}^* C_{k'j}C_{kj'}^* C_{k'j'}] \abs{z_j(0)}^2\abs{z_{j'}(0)}^2 = \sum_j \E[\abs{C_{kj}}^2\abs{C_{k'j}}^2] \abs{z_j(0)}^4 \approx \frac{1}{N^2}\sum_j\abs{z_j(0)}^4. 
\end{equation}
This means that for large $N$ we can approximately replace $\sum_j C_{kj}^* C_{k'j}\abs{z_j(0)}^2$ with the fluctuation $\frac{1}{N}\sqrt{\sum_j\abs{z_j(0)}^4}$. Hence, 
\begin{equation}
 \sum_{k,k':k\neq k'}\mu_{kk} \mu_{k'k'}   \sum_{j,r:j\neq r}C_{kj}^* C_{k'j} \abs{z_j(0)}^2 C_{k'r}^* C_{kr}\abs{z_r(0)}^2 \approx \left(\frac{N^*}{N}\right)^2 \omega_\text{av}^2 \sum_j \abs{z_j(0)}^4 \ll \overline{A_M}_\infty^2.
\end{equation}
In conclusion, the RHS of Eq.~\eqref{eqn:varAM} is dominated by $\overline{A_M}_\infty^2$, and we obtain the result in Eq.~\eqref{eqn:fluctuationsAM}.
}

\section{Relation between Khinchin's approach and the results in this work}

\mar{
In his treatise on the mathematical foundations of statistical mechanics \cite{Khinchin1949}, Khinchin showed the validity of statistical mechanics in classical systems withouth ergodic theory. He only dealt with non-interacting systems with Hamiltonian
\begin{equation}
    H = \sum_{j=1}^N H_j,
\end{equation}
and observables written as a sum of many single-particle terms,
\begin{equation}
\label{eqn:khinchinS_observable}
    f = \sum_{j=1}^Nf_j.
\end{equation}
For instance, he proposed as an example a gas of non-interacting molecules. His results were later extended to weakly interacting systems by Mazur and van der Linden \cite{Mazur1963}.
}

\mar{Khinchin's idea is based on the fact that $f$ in Eq.~\eqref{eqn:khinchinS_observable} is such that \cite{castiglione2008chaos,Baldovin2024found}
\begin{equation} \label{eqn:khinchin_var}
    \langle f \rangle = \mathcal{O}(N), \quad \sigma^2_f =\langle (f-\langle f \rangle )^2\rangle = \mathcal{O}(N),
\end{equation}
where $\langle \cdot \rangle$ indicates the microcanonical ensemble average. This is a fundamental property for statistical mechanics, which, for instance, is emphasized also in the introductory chapter of Landau and Lifshitz's celebrated treatise on statistical physics \cite{landau}. There, it is also presented as the fundamental reason why statistical mechanics, at least in its standard formulation, works for large systems only, $N\gg 1$. Khinchin's idea is similar. Here, we follow the derivation presented in \cite{Baldovin2024found}. Using Markov's inequality, from Eq.~\eqref{eqn:khinchin_var} we observe that the measure of the set of points in the microcanonical ensemble for which $\abs{f-\langle f\rangle}/\abs{\langle f \rangle}\geq a$, for any $a>0$, is bounded by the standard deviation of $f$. In terms of probability with respect to the microcanonical distribution, 
\begin{equation}\label{eqn:prob_nontimeav}
    \text{Prob}\left(\frac{\abs{f-\langle f \rangle}}{\abs{\langle f \rangle}}\geq a\right)\leq \frac{\sigma}{a \abs{\langle f \rangle}}. 
\end{equation}
Moreover, we can use the intuitive fact that $\langle (\bar{f}_\infty-\langle f\rangle)^2\rangle \leq \langle (f-\langle f \rangle )^2\rangle  $, where $\bar{f}_\infty$ is the long-time average in Eq.~\eqref{eqn:timeAv}. Then, through the Cauchy-Schwartz inequality the same bound can be written for the long-time average of $f$:
\begin{equation}\label{eqn:prob_timeav}
    \text{Prob}\left(\frac{\abs{\bar{f}_\infty-\langle f \rangle}}{\abs{\langle f \rangle}}\geq a\right)\leq \frac{\sigma}{a \abs{\langle f \rangle}}. 
\end{equation}
Next, if we choose, for instance, $a= N^{-1/4}$, the probabilities in Eqs.~\eqref{eqn:prob_nontimeav} and~\eqref{eqn:prob_timeav} vanish for large $N$. In other words, the measure of the set of points in the microcanonical ensemble for which the value of $f$, or its long-time average $\bar{f}_\infty$, deviates from the microcanonical average $\langle f \rangle$ more than a quantity that goes as $N^{-1/4}$ (increasingly small for large $N$), vanishes for $N\rightarrow \infty$: for almost all points in the microcanonical ensemble, $f$ is already at thermal equilibrium, independently of the time evolution.}

\mar{Khinchin explicitly remarks that the physical assumption underlying this reasoning is that initial conditions (i.e., states) in the microcanonical ensemble that deviate from the thermal equilibrium value of $f$ ``appear very infrequently in practice" (\cite{Khinchin1949}, section 13), as they form a zero-measure subset within the ensemble. While one may engineer a system with an initial state far from equilibrium, such a setup would be an ad hoc realization of a highly atypical state, rarely encountered in reality. Whether this description adequately captures the natural world---beyond what physicists may study and manipulate in the laboratory---is ultimately a matter of interpretation. Here, we do not want to get into the philosophical foundations of statistical mechanics.  
}

\mar{Let us now discuss the relation of our results with Khinchin's idea. First of all, in this work we deal with strongly interacting systems. We can write the Hamiltonian as a separable one in the basis of the normal modes (Eq.~\eqref{eqn:DiagHamNormalModes_first}), but the observables we are studying cannot be simply written as in Eq.~\eqref{eqn:khinchinS_observable}. This is true for both the microscopic observables (e.g., the energies of the rotated modes) and the macroscopic ones (sum of many single-particle energies). In fact, the rotated modes in Eq.~\eqref{eqn:HamRandomRotRotModes} or the local modes of the particles in Eq.~\eqref{eqn:randomHam} can be strongly interacting. As such, Khinchin's results do not directly apply to our models, and our study is just inspired by them.
}

\mar{What we have proven, essentially, is the fact that the energies of the rotated or local modes, which now we indicate as $f$ for simplicity, assuming that the rotation is random and analogously for the random matrix in Eq.~\eqref{eqn:randomHam}, satisfy the relation
\begin{equation} \label{eqn:var_obs_khinchinquasi}
    \frac{\langle (\bar{f}_\infty-\langle f\rangle)^2\rangle}{\langle f \rangle^2}=\mathcal{O}(1/N).
\end{equation}
This property resembles Eq.~\eqref{eqn:khinchin_var}, but with the long-time average of the observable instead of $f$.  Eq.~\eqref{eqn:var_obs_khinchinquasi} can be immediately employed to prove the typicality of thermalization for the long-time average of the single-particle rotated energies, now expressed in terms of Eq.~\eqref{eqn:prob_timeav}. The fact that for classical microscopic observables we obtain Eq.~\eqref{eqn:var_obs_khinchinquasi} instead of Eq.~\eqref{eqn:khinchin_var} is quite natural: classical microscopic observables fluctuate in time even at equilibrium, as we have discussed in Appendix B. Eq.~\eqref{eqn:khinchin_var} was proposed by Khinchin (or analogously, in a different context, by Landau and Lifshitz \cite{landau}) thinking of macroscopic observables. In fact, in the Appendix B we have also proven that the classical macroscopic observables satisfy Eq.~\eqref{eqn:khinchin_var}, and consequently Eq.~\eqref{eqn:prob_nontimeav}. If we pick a random initial condition from the microcanonical ensemble, for $N\gg1$ the macroscopic observable is thermalized with probability approaching 1, at any time. For microscopic quantum observables in the case of fermions or hard-core bosons, the long-time variance also vanishes, recovering Eq.~\eqref{eqn:khinchin_var}. The fact that the model is integrable does not play any role for this type of thermalization to appear. Last but not least, we have also proven that Eqs.~\eqref{eqn:khinchin_var} and~\eqref{eqn:var_obs_khinchinquasi} in our model hold not only in the microcanonical ensemble, but also in the restricted microcanonical ensemble where we excited only a fraction $N^*\ll N$ of the normal modes, provided $N^*\gg1$. Finally, note that the expression in Eqs.~\eqref{eqn:khinchin_var} and~\eqref{eqn:var_obs_khinchinquasi} is conceptually different from some known results about the mixing properties of classical \cite{Lanford1975} and quantum \cite{Graffi1996} infinite harmonic crystals (see also \cite{Jona-Lasinio1996} for extensions to non-integrable systems), which do not consider the variance with respect to the microcanonical ensemble.
}

\mar{Let us now consider the case of roughly equal initial populations, for which we reach energy equipartition at equilibrium without thermalization. Note that this scenario is again fully captured by Eqs.~\eqref{eqn:khinchin_var},~\eqref{eqn:prob_nontimeav},~\eqref{eqn:prob_timeav} and~\eqref{eqn:var_obs_khinchinquasi}, as we are dealing with the same observables satisfying these relations. By choosing roughly equal initial populations, we are selecting some states in a zero-measure subset of the microcanonical ensemble, as explained in Appendix A, which according to Khinchin's should appear very infrequently in practice. We are therefore choosing an ``atypical state'' that is far from equilibrium.}

\mar{To conclude, we have rigorously proven that Khinchin's idea, formalized by Eqs.~\eqref{eqn:khinchin_var},~\eqref{eqn:prob_nontimeav},~\eqref{eqn:prob_timeav} and~\eqref{eqn:var_obs_khinchinquasi}, is fruitful even in scenarios that go beyond Khinchin's original proposal, including, quite remarkably, quantum mechanics. Moreover, we have proven Eqs.~\eqref{eqn:var_obs_khinchinquasi} and~\eqref{eqn:khinchin_var} for a very general class of systems, namely harmonic systems, that include several paradigmatic models in both classical and quantum physics. Next, we have shown that, within this class of systems, these relations hold not only in the microcanonical ensemble, but also on a restricted version thereof, where we excite only a fraction $N^*$ of modes, with $1\ll N^*\ll N$. In addition, we have done so by looking at the dynamics of the observables of interest, while Khinchin's results are based on the structure of the microscopical ensemble  and do not take into account the time evolution at all. Taking the dynamics into account allows us, for instance, to clearly tell the difference between the different choices of initial conditions we have considered in our work (see the results about the time variance in the previous section and the example in Appendix B). Finally, our results are fully analytical, and do not just rely on numerical simulations.}

\section{A physically relevant case: random classical Hamiltonian without position-momentum coupling}
\label{sec:addRes}

Let us consider a generic (classical) Hamiltonian of a collection of $N$ particles with equal mass $m=1$ immersed in a harmonic potential, given by
\begin{equation}
    H = \sum_j \frac{p_j^2}{2} + \sum_{\substack{j,k=1}}\frac{v_{jk}}{2} q_jq_k=\frac{1}{2}\mathbf{p}^T \mathsf{K} \mathbf{p}+\frac{1}{2}\mathbf{q}^T \mathsf{V}\mathbf{q},
\end{equation}
where we have introduced the kinetic matrix $\mathsf{K}=\mathbb{I}$ and the real potential matrix $\mathsf{V}=\mathsf{V}^T$. For instance, in the paradigmatic case of the linear chain with fixed boundary conditions, we set $v_{jk}=-\delta_{j,k\pm1}$ and $v_{jj}=2$ \cite{Cocciaglia2022}. 

Let us now introduce the real orthogonal matrix of eigenvectors $C$, with $C^T C =\mathbb{I}$. We introduce the \textit{normal modes} as
\begin{equation}
    \mathbf{Q} = C^T \mathbf{q}, \quad \mathbf{P}= C^T \mathbf{p}.
\end{equation}
Then, $\{Q_j,P_k\}=\delta_{jk}$, and
\begin{equation}
    H = \frac{1}{2}\mathbf{P}^T \mathsf{K} \mathbf{P}+\frac{1}{2}\mathbf{Q}^T \mathsf{D}\mathbf{Q},
\end{equation}
where $\mathsf{D}=C^T \mathsf{V} C = \text{diag}(\omega_j)$. The Hamiltonian is then decomposed into a sum of non-interacting normal modes,
\begin{equation}
\label{eqn:HamFree}
    H = \sum_j \frac{1}{2}(\omega_j^2 Q_j^2 + P_j^2).
\end{equation}
Each normal mode oscillates with the corresponding frequency $\omega_j$. More in particular, we have
\begin{equation}
\label{eqn:solutionLinear}
    Q_j(t) = A_j \cos(\omega_j t +\phi_j),\qquad P_j(t) = -\omega_j A_j \sin(\omega_j t +\phi_j),
\end{equation}
with the initial conditions determining the amplitude and the phase of the oscillations:
\begin{equation}
\label{eqn:initialCond}
    A_j= \sqrt{Q_j(0)^2+(P_j(0)/\omega_j)^2},\qquad \phi_j=\arctan\left(-\frac{P_j(0)}{\omega_j Q_j(0)}\right).
\end{equation}
For the linear chain with fixed boundary conditions, we find the eigenfrequencies
\begin{equation}
    \label{eqn:eig1Dchain}
    \omega_j = 2\sin\left(\frac{\pi j}{2N+2}\right),
\end{equation}
and the eigenvectors whose components are given by
\begin{equation}
    \label{eqn:eigVec1Dchain}
    C_{jk}=\sqrt{\frac{2}{N+1}}\sin\left(\frac{\pi j k}{N+1}\right).
\end{equation}

Let us now compute the time average of the local energies $\tilde{E}_j^{(loc)}=\frac{p_j^2}{2}+\frac{v_{jj}}{2}q_j^2$, analogously to Eq.~\eqref{eqn:Energy_Loc_modes}:
\begin{equation}
\label{eqn:solutionFinal}
    \begin{split}
        O_j^{(loc)}&=\overline{\tilde{E}_j^{(loc)}}_t :=\frac{1}{t}\int_0^t ds \tilde{E}_j(s)=\frac{1}{t}\int_0^t ds\frac{p_j^2(s)+v_{jj}q_j^2(s)}{2}= \frac{1}{t}\int_0^t ds\left[(\sum_k C_{jk} P_k(s))^2+v_{jj}(\sum_k C_{jk} Q_k(s))^2\right]\\
        & = \frac{1}{t}\int_0^t ds \left[\sum_k\sum_{k'=1}C_{jk}C_{jk'}(  P_k(s)P_{k'}(s)+v_{jj} Q_k(s)Q_{k'}(s))\right]\\
        & = \sum_{\substack{k,k'=1 }} \frac{C_{jk}C_{jk'}A_k A_{k'}}{t}\int_0^t ds (  \omega_k\omega_{k'}\sin(\omega_k s+\phi_k)\sin(\omega_{k'}s+\phi_{k'})+v_{jj} \cos(\omega_k s+\phi_k)\cos(\omega_{k'}s+\phi_{k'}))\\
         & = \sum_{\substack{k,k'=1 \\ k\neq k'}} \frac{C_{jk}C_{jk'}A_k A_{k'}}{2t}\left[ \omega_k\omega_{k'}\left(\frac{\sin((\omega_k-\omega_{k'})t+\phi_k-\phi_{k'})-\sin(\phi_k-\phi_{k'})}{\omega_k-\omega_{k'}}\right.\right.\\
         &\quad\left.-\frac{\sin((\omega_k+\omega_{k'})t+\phi_k+\phi_{k'})-\sin(\phi_k+\phi_{k'})}{\omega_k+\omega_{k'}}\right)+v_{jj} \left(\frac{\sin((\omega_k-\omega_{k'})t+\phi_k-\phi_{k'})-\sin(\phi_k-\phi_{k'})}{\omega_k-\omega_{k'}}\right.\\
         &\left.\left.\quad+\frac{\sin((\omega_k+\omega_{k'})t+\phi_k+\phi_{k'})-\sin(\phi_k+\phi_{k'})}{\omega_k+\omega_{k'}}\right)\right]\\
         &\quad +\sum_k \frac{C_{jk}^2 A_k^2}{4t\omega_k}\left[\omega_k^2 \left(2\omega_k t-\sin 2\phi_k+\sin(2(\omega_k t+\phi_k))\right)+v_{jj}\left(2\omega_k t+\sin 2\phi_k -\sin(2(\omega_k t+\phi_k))\right)\right].\\
    \end{split}
\end{equation}
We have used the analytical solution of the dynamics of the normal modes in Eq.~\eqref{eqn:solutionLinear} and the amplitudes and phases given by the initial conditions according to Eq.~\eqref{eqn:initialCond}.

In the limit of $t\rightarrow\infty$, all the constant and oscillatory terms in Eq.~\eqref{eqn:solutionFinal} disappear due to the factor $1/t$. Therefore, only the linear terms survive, and the time average at infinite time simply reads
\begin{equation}
\label{eqn:infiniteDistribution_lapl}
    \overline{\tilde{E}_j^{(loc)}}_{\infty}=\overline{\tilde{E}_j^{(kin)}}_{\infty}+\overline{\tilde{E}_j^{(pot)}}_{\infty}=\sum_k \frac{C_{jk}^2 A_k^2}{2}(\omega_k^2+v_{jj}).
\end{equation}
We have divided the final expression into kinetic $\overline{\tilde{E}_j^{(kin)}}_{\infty}=\sum_k\frac{C_{jk}^2 A_k^2}{2}\omega_k^2$ and potential energy $\overline{\tilde{E}_j^{(pot)}}_{\infty}=\sum_k\frac{C_{jk}^2 A_k^2}{2}v_{jj}$.
The phase of the oscillations does not matter at all, as we may expect. Eq.~\eqref{eqn:infiniteDistribution_lapl} is the main quantity we have to evaluate and, once again, we note that it crucially depends on the elements of the eigenvectors $C_{jk}$.

Eq.~\eqref{eqn:infiniteDistribution_lapl} is decomposed into a kinetic and a potential contribution. If we assume that $C_{jk}^2$ are delocalized for large $N$, equipartition follows immediately from the proof for a random matrix $M$ in the previous section. For the potential part, we only need to assume the standard condition for $p_{0,j}=A_j^2$. For the kinetic part, we also need to assume the standard condition for the frequencies as done in the case of randomly rotated modes. 

\begin{figure*}[h!]

    \centering
    \includegraphics[scale=0.48]{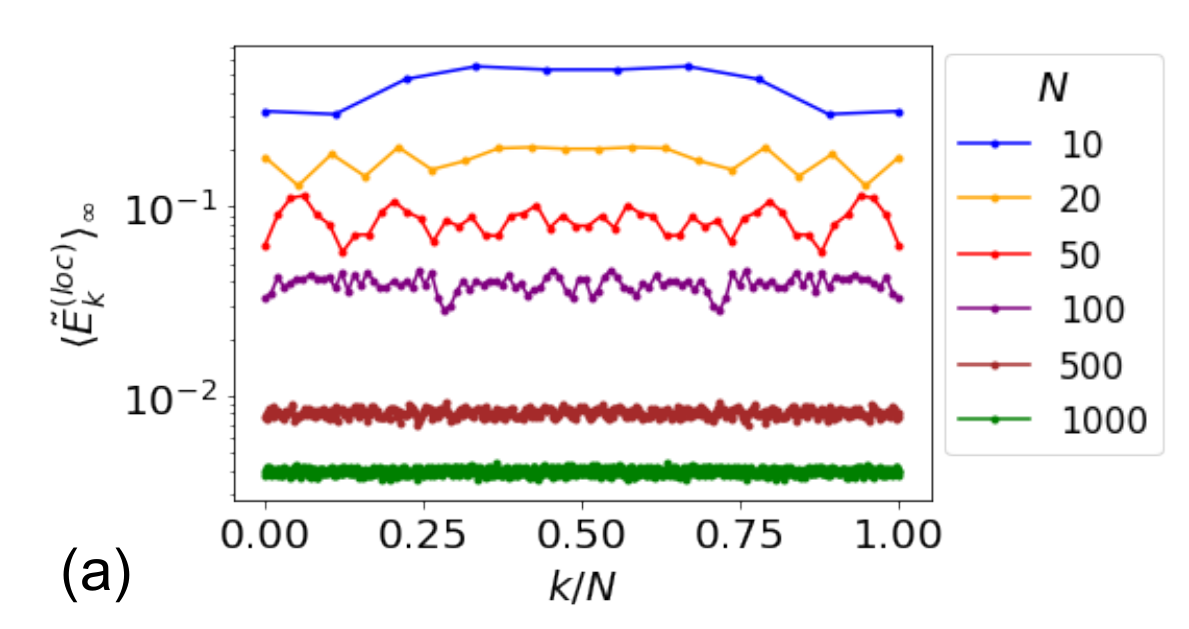}
    \includegraphics[scale=0.48]{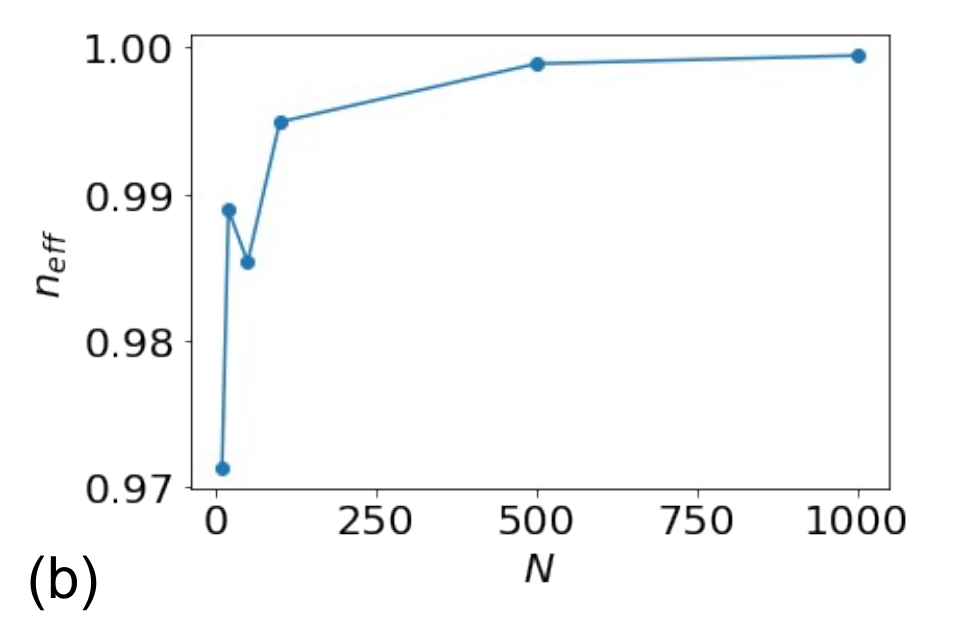}
    \caption{Numerical results for the energy of the local particles of the linear chain according to Eq.~\eqref{eqn:infiniteDistribution_lapl}, where the deterministic eigenvectors of the linear chain are given by Eq.~\eqref{eqn:eigVec1Dchain}. The initial populations are randomly distributed among the normal modes. (a): Distribution of ${O}^{(loc)}_k$ over $k$. (b): $n_\text{eff}$ (Eq.~\eqref{eqn:neff}) as a function of $N$ for the scenario in (a). Energy equipartition cleary emerges in the limit of large $N$.}
    \label{fig:linearChain}
\end{figure*}

Let us now show the emergence of equipartition in the case of the linear chain, whose eigenvectors are given by Eq.~\eqref{eqn:eigVec1Dchain}. As the model is fixed and deterministic, the matrix of eigenvectors is not random anymore. This being said, having a look at 
\begin{equation}
C_{jk}^2 = \frac{2}{N+1}\sin^2\left(\frac{\pi j k}{N+1}\right),    
\end{equation}
in the limit of $N\gg1$ we can suppose that the argument of the $\sin^2$ function is extracted randomly from $[0,2\pi)$ with weight $1/2\pi$ \cite{Baldovin2023}. Then, we can easily compute the moments of $C_{jk}^2$ treated as a random variable:
\begin{equation}
    \E[C_{jk}^2] = \frac{1}{N+1},\qquad \E[C_{jk}^4]= \frac{3}{2(N+1)^2}. 
\end{equation}
Using these moments we immediately prove that the ``mean value'' of the energies of the local modes of the linear chain gets equipartitioned with vanishing fluctuations for $N\gg1$, following the same steps as in the previous subsection. 
A numerical validation of this result is shown in Fig.~\ref{fig:linearChain}.%SM.

\end{document}